\documentclass[12pt]{article} 
\usepackage{amssymb}
\renewcommand{\Bbb}{\bf}
\newcommand{\newsubsection}[1]{
\vspace{1cm}
\pagebreak[3]
\addtocounter{subsection}{1}
\addcontentsline{toc}{subsection}{\protect
\numberline{\arabic{section}.\arabic{subsection}}{#1}}
\noindent{\it \thesubsection.  #1}                 
\nopagebreak
\vspace{2mm}
\nopagebreak}
\newcommand{\newsection}[1]{
\vspace{15mm}
\pagebreak[3]
\addtocounter{section}{1}
\setcounter{equation}{0}
\setcounter{subsection}{0}
\setcounter{footnote}{0}
\addcontentsline{toc}{section}{\protect
\numberline{\arabic{section}}{{\rm #1}}}
\noindent{\bf \thesection.  #1}                 
\nopagebreak
\vspace{2mm}
\nopagebreak}
\renewcommand{\theequation}{\thesection.\arabic{equation}}

\newlength{\extraspace}
\newlength{\extraspaces}
\newcounter{dummy}
\newcommand{\be}{\begin{equation}
\addtolength{\abovedisplayskip}{\extraspaces}
\addtolength{\belowdisplayskip}{\extraspaces}
\addtolength{\abovedisplayshortskip}{\extraspace}
\addtolength{\belowdisplayshortskip}{\extraspace}}
\newcommand{\ee}{\end{equation}}
\newcommand{\ba}{\begin{eqnarray}
\addtolength{\abovedisplayskip}{\extraspaces}
\addtolength{\belowdisplayskip}{\extraspaces}
\addtolength{\abovedisplayshortskip}{\extraspace}
\addtolength{\belowdisplayshortskip}{\extraspace}}
\newcommand{\ea}{\end{eqnarray}}
\newcommand{\nonu}{\nonumber \\[2mm]}
\newcommand{\is}{& \!\! = \!\! &}
\newcommand{\ban}{\begin{eqnarray*}
\addtolength{\abovedisplayskip}{\extraspaces}
\addtolength{\belowdisplayskip}{\extraspaces}
\addtolength{\abovedisplayshortskip}{\extraspace}
\addtolength{\belowdisplayshortskip}{\extraspace}}
\newcommand{\ean}{\end{eqnarray*}}
\newcommand{\baa}{                         
\addtocounter{equation}{1}
\setcounter{dummy}{\value{equation}}
\setcounter{equation}{0}
\renewcommand{\theequation}{\thesection.\arabic{dummy}\alph{equation}}
\begin{eqnarray}
\addtolength{\abovedisplayskip}{\extraspaces}
\addtolength{\belowdisplayskip}{\extraspaces}
\addtolength{\abovedisplayshortskip}{\extraspace}
\addtolength{\belowdisplayshortskip}{\extraspace}}
\newcommand{\eaa}{                                       
\end{eqnarray}
\setcounter{equation}{\value{dummy}}
\renewcommand{\theequation}{\thesection.\arabic{equation}}}

\input epsf
\newcounter{fignum}
\newcommand{\figuurnum}{\arabic{fignum}}

\newcommand{\figuurplus}[3]{
\addtocounter{fignum}{1}
\addcontentsline{lof}{figure}{\protect
\numberline{\arabic{section}.\arabic{fignum}}{#3}}
\hspace{-3mm}{\it fig.}\ \figuurnum.
\begin{figure}[t]\begin{center}
\leavevmode\hbox{\epsfxsize=#2 \epsffile{#1.eps}}\\[3mm]
\parbox{10cm}{\small \bf Fig.\ \figuurnum: \it #3}
\end{center} \end{figure}\hspace{-1.5mm}}

\newcounter{tabnum}
\newcounter{xxx}
\newcommand{\bl}{\begin{list}{({\it\roman{xxx}})}{\usecounter{xxx}}}
\newcommand{\el}{\end{list}}
\renewcommand{\d}{{{\partial}}}

\newcommand{\ppt}[1]{{\partial \over \partial t}}            
\newcommand{\ppx}[1]{{\partial \over \partial x}}            
\newcommand{\pqt}[1]{{\partial^2 \over \partial t^2}}            
\newcommand{\pqx}[1]{{\partial^2  \over \partial x^2}}            
\newcommand{\twomatrixd}[4]{{\left(\begin{array}{cc}
\displaystyle #1 & \displaystyle #2\\[2mm]
\displaystyle  #3  & \displaystyle #4 \end{array}\right)}}

\newcommand{\ie}{{\it i.e.}}

\renewcommand{\l}{\langle}

\renewcommand{\.}{\cdot}
\renewcommand{\ll}{{\lambda}}

\newcommand{\half}{{\textstyle{1\over 2}}}

\newcommand{\Z}{{\Bbb Z}}
\newcommand{\R}{{\Bbb R}}

\newcommand{\C}{{\Bbb C}}

\renewcommand{\H}{{\Bbb H}}

\newcommand{\cO}{{\cal O }}            
\newcommand{\cH}{{\cal H }}

\newcommand{\ra}{\rightarrow}

\newcommand{\opda}{\mathop{\downarrow}}

\newcommand{\inv}{^{-1}}
\newcommand{\V}{{\cal V}}

\newcommand{\Tr}{{\rm Tr}\,}

\newcommand{\dbar}{{\overline{\partial}}}

\newcommand{\pbar}{{\overline{p}}}
\newcommand{\qbar}{{\overline{q}}}

\newcommand{\ybar}{{\overline{y}}}

\newcommand{\Lbar}{{\overline{L}}}

\newcommand{\thetabar}{{\overline{\theta}}}

\newcommand{\cM}{{\cal M}}
\newcommand{\cF}{{\cal F}}
\newcommand{\cN}{{\cal N}}
\newcommand{\cQ}{{\cal Q}}

\renewcommand{\Im}{{\rm Im\,}}

\newcommand{\option}[4]{ \left\{ \begin{array}{ll}
#1, & \hbox{#2}, \\[2mm]
#3, & \hbox{#4}. \end{array} \right. }

\newcommand{\ext}{{\raisebox{.2ex}{$\textstyle \bigwedge$}}}

\def\a{\alpha} 
 
\def\g{\gamma} 
\def\G{\Gamma}
\def\e{\epsilon}

\def\th{\theta}

\def\l{\lambda} 
\def\L{\Lambda} 
\def\m{\mu}
\def\n{\nu}
 
\def\s{\sigma} 
\def\t{\tau}
\def\f{\phi} 
\def\F{\Phi} 
\def\w{\omega}
\def\W{\Omega}

\def\<{\langle}
\def\>{\rangle}

\newcommand{\id}{{\bf 1}}

\newfont{\gothic}{eufm10 scaled\magstep1}

\newcommand{\commutingdiagram}[8]{
\begin{array}{ccc}
 \displaystyle #1 & \displaystyle \mathop{\longrightarrow}^{#2} & 
 \displaystyle #3 \\[3mm]
\opda {\scriptstyle #4} & & \opda {\scriptstyle #5}\\[3mm]
 \displaystyle #6 &  \displaystyle\mathop{\longrightarrow}^{#7} & 
 \displaystyle #8
\end{array}}

\newcommand{\sdim}{{\rm sdim}\,}
\newcommand{\sdet}{{\rm sdet}}
\newcommand{\str}{{\rm sTr}\,}
\newcommand{\cL}{{\cal L}}

\newcommand{\Hilb}{{\rm Hilb}}
\renewcommand{\bar}{\overline}
\newcommand{\SNX}{{S^N\!X}}
\newcommand{\twosub}[2]{_{\scriptstyle #1 \atop\strut\scriptstyle #2}}
\newcommand{\boxx}[2]
{\; {\hbox{\small
$#1$}\,\mathop{\fbox{\rule[-0.2mm]{0mm}{2mm}\hspace{2mm}}}_{\hbox{\small
$#2$}}}}
\newcommand{\nn}{^{(n)}}

\begin{document}
\thispagestyle{empty}

\addtolength{\baselineskip}{.5mm}
\begin{flushright}
October 1998\\
{\sc hep-th/9912104}\\
\end{flushright}

\vspace{8mm}

\begin{center}
{\Large \sc Fields,~Strings,~Matrices~and~Symmetric~Products\footnote{Based 
on lectures given among others
at the {\it Geometry and Duality Workshop} at the Institute for Theoretical
Physics, UC Santa Barbara, January 1998 and the {\it 
Spring School on String Theory and Mathematics,}
Harvard University, May 1998. \\ Published in {\sl Moduli
of Curves and Abelian Varieties, The Dutch Intercity Seminar on
Moduli,} C. Faber and E. Looijenga (Eds.), Vieweg, Aspects of
Mathematics E33, 1999.}}\\[35mm]
\addtocounter{footnote}{1}
{\sc Robbert Dijkgraaf}\\[5mm]
{\it Departments of Mathematics and Physics\\
University of Amsterdam\\
Plantage Muidergracht 24, 1018 TV Amsterdam}\\[3mm]
{\tt rhd@wins.uva.nl}\\[2cm] 
{\sc Abstract}
\end{center}

\begin{quote} 

\noindent In these notes we review the role played by the quantum mechanics
and sigma models of symmetric product spaces in the light-cone
quantization of quantum field theories, string theory and matrix theory.
\end{quote}

\newpage

\newsection{Introduction}

For more than a decade now, string theory has been a significant,
continuous influence in mathematics, ranging from fields as diverse as
algebraic geometry to representation theory. However, it is fair to say
that most of these applications concerned the so-called first-quantized
formulation of the theory, the formulation that is used to describe to
propagation of a single string. In contrast with point-particle
theories, in string theory the first-quantized theory is so powerful
because it naturally can be extended to also describe the perturbative
interactions of splitting and joining of strings by means of Riemann
surfaces of general topology. Study of these perturbative strings has
led to series of remarkable mathematical developments, such as
representation theory of infinite-dimensional Lie algebras, mirror
symmetry, quantum cohomology and Gromov-Witten theory. 

The second-quantized formalism, what is sometimes also refered to as
string field theory, has left a much smaller mathematical imprint.
Although there is a beautiful geometrical and algebraic structure of
perturbative closed string field theory, developed mainly by Zwiebach
\cite{zwiebach}, which is built on deep features of the moduli space of
Riemann surfaces, it is very difficult to analyze, perhaps because it is
intrinsically perturbative. Yet, in recent years it has become clear
recently that the non-perturbative mathematical structure of string
theory is even richer than the perturbative one, with even bigger
symmetry groups---the mysterious $U$-duality groups
\cite{hull-townsend}. The appearance of D-branes \cite{polch} and a
eleven-dimensional origin in the form of M-theory \cite{witten-M} can
only be properly understood from a second-quantized point of view.

At present there is only one candidate for a fundamental description of
non-perturbative string theory, which is matrix theory \cite{bfss}. In
matrix theory an important role is played by non-abelian gauge fields,
and the strings and conformal field theory only emerge in a certain weak
coupling limit. We will not review much of matrix theory in these notes
but refer to for example \cite{bilal,banks,susskind,maxmic}. Important
is that matrix theory makes directly contact with the second-quantized
theory, indeed Fock spaces are one of the ubiquitous ingredients, and
much of the notes will focus on this correspondence, also reviewing the
work of \cite{ell-genus,matrix-string}.

\newsubsection{Hamilton vs Lagrange: 
representation theory and automorphic forms}

One of the most remarkable insights provided by string theory, or more
properly conformal field theory, is the natural explanation it offers
of the modular properties of the characters of affine Kac-Moody
algebras, Heisenberg algebras, and other infinite-dimensional Lie
algebras. At the hart of this explanation---and in fact of much of the
applications of field theory in mathematics---lies the equivalence
between the Hamiltonian and Lagrangian formulation of quantum
mechanics and quantum field theory.

This equivalence roughly proceeds as follows
(see also \cite{witten-icm}).  In the Hamiltonian
formulation one considers the quantization of a two-dimensional
conformal field theory on a space-time cylinder $\R \times S^1$.  The
basic object is the loop space $\cL X$ of maps $S^1 \ra X$ for some
appropriate target space $X$. The infinite-dimensional Hilbert space
$\cH$ that forms the representation of the algebra of quantum
observables is then typically obtained by quantizing the loop space
$\cL X$. This Hilbert space carries an obvious $S^1$ action, generated
by the momentum operator $P$ that rotates the loop, and the character 
of the representation is defined as
\be
\chi(q)= \Tr _{\strut \! \cH} q^{P}
\ee
with $q=e^{2\pi i \tau}$ and $\t$ in the complex upper
half-plane $\H$. The claim is that these characters are always some kind
of modular forms. From the representation theoretic point of view it is not
at all clear why there should be a natural action of the modular group
$SL(2,\Z)$ acting on $\tau$ by linear fractional transformations. In
particular the transformation $\t \ra -1/\t$ is rather mysterious.

In the Lagrangian formulation, however, the character $\chi(q)$, or more
properly the partition function, is computed by considering the quantum
field theory on a Riemann surface with topology of a two-torus $T^2=S^1
\times S^1$, \ie, an elliptic curve with modulus $\t$. The starting
point is the path-integral over all maps $T^2 \ra X$. Since we work with
an elliptic curve, the modularity is built in from the start. The
transformation $\t \ra -1/\t$ simply interchanges the two $S^1$'s.
Changing from the Hamiltonian to the Lagrangian perspective, we
understand the appearance of the modular group $SL(2,\Z)$ as the
`classical' automorphism group of the two-torus. This torus is obtained
by gluing the two ends of the cylinder $S^1 \times \R$, which is the
geometric equivalent of taking the trace. Note that in string theory
this two-torus typically plays the role of a {\it world-sheet}.

In second-quantized string theory we expect a huge generalization of this
familiar two-dimensional story.  The operator algebras will be much
bigger (typically, generalized Kac-Moody algebras) and also the
automorphism groups will not be of a classical form, but will reflect
the `stringy' geometry at work.  An example we will discuss in great
detail in these notes is the quantization of strings on a space-time
manifold of the form 
\be 
M = \R \times S^1 \times X,
\ee
with $X$ a compact simply-connected Riemannian manifold.  Quantization
leads again to a Hilbert space $\cH$, but this space carries now at
least two circle actions. 

First, we have again a momentum operator $P$ that generates the
translations along the $S^1$ factor. Second, there is also a winding
number operator $W$ that counts how many times a string is wound
around this circle. It labels the connected components of the loop
space $\cL M$. A state in $\cH$ with eigenvalue $W=m\in
\Z$ represents a string that is wound $m$ times around the $S^1$.
In this way we can define a two-parameter character
\be
\chi(q,p) = \Tr_{\strut \! \cH} p^W q^{P},
\label{char}
\ee
with $p= e^{2\pi i \s}$, $q=e^{2\pi i \tau}$, with both $\s,\t\in\H$.
We will see in concrete examples that these kind of expressions will
be typically the character of a generalized Kac-Moody algebra and
transform as automorphic forms.

The automorphic properties of such characters become evident by changing
again to a Lagrangian point of view and computing the partition function on
the compact manifold $T^2 \times X$. Concentrating on the $T^2$
factor, which now has an interpretation as a {\it space-time,}
the string partition function carries a manifest $T$-duality
symmetry group
\be
SO(2,2,\Z) \cong SL(2,\Z) \times SL(2,\Z),
\ee
which is the `stringy' automorphism group of $T^2$.

Let us explain briefly how this group acts on the moduli $\s,\t$. Since
the string theory is not a conformal field theory, the partition
function will depend both on the modulus $\t$ of $T^2$ and on its volume
$g$. Furthermore there is an extra dependence on a constant 2-form field
$\th\in H^2(T^2,\R/\Z)$. These two extra data are combined in a second
complex `modulus' $\s=\th + ig$. The $T$-duality group $SO(2,2,\Z)$ will
now acts on the pair $(\s,\t)$ by separate fractional linear
transformations and the generalized character (\ref{char}) will be some
automorphic form for this group. Of course only the second $SL(2,\Z)$
factor has a clear geometric interpretation. The first factor, that
exchanges large and small volume $\s \ra -1/\s$, as a complete stringy
origin.

The appearance of the $T$-duality group $SO(2,2,\Z)$ as a symmetry group
of the two-torus is most simply
explained by considering a single string. We are then dealing with
the loop space $\cL T^2$. If the torus is given by the quotient
$\R^2/\L$, with $\L$ a two-dimensional lattice, the momenta of such
a string take value in the dual lattice
\be
P = \oint \dot x \in \L^*.
\ee
The winding numbers, that label the components of $\cL T^2$, lie
in the original lattice
\be
W = \oint dx \in \L.
\ee
Therefore the total vector $v=(W,P)$ can be seen as an element
of the rank 4, signature (2,2), even, self-dual Narain lattice
\be
v=(W,P)\in \G^{2,2} = \L \oplus \L^*,\qquad
v^2 = 2 \, W\cdot P
\ee
The $T$-duality group appears now as the automorphism group of the lattice
$\G^{2,2}$. In the particular example we will discuss in detail,
where the manifold $X$ is a Calabi-Yau space, there will be an
extra quantum number and the lattice will be enlarged to a
signature (2,3) lattice. Correspondingly, the automorphism group
will be given by $SO(3,2,\Z) \cong Sp(4,\Z)$.

\newsection{Particles, symmetric products and fields}

It is well-known wise-crack that first-quantization is a mystery but
second-quantization a functor. Indeed, for a free theory second
quantization involves nothing more than taking symmetric products. We
obtain the second-quantized Hilbert space from the first-quantized
Hilbert space $\cH$ as the free symmetric algebra $S\cH$. Yet, recent
developments in string theory (and in certain field theories that are
naturally obtained as limits of string theories) have provided us with a
fresh outlook on this familiar subject. In particular this new approach
allow us to include interactions in new ways.

\newsubsection{Second-quantization of superparticles}

Let us start by considering a well-known case: a point-particle moving
on a compact oriented Riemannian manifold $X$. The first-quantization
`functor' $\cQ^1$ of quantum mechanics
assigns to each manifold $X$ a Hilbert space $\cH$
and a Hamiltonian $H$,
\be
\cQ^1: X \mapsto (\cH,H).
\ee
As is well-know, in (bosonic) quantum mechanics the Hilbert space is
given by the square-integrable functions on $X$, $\cH = L^2(X)$,
together with the positive-definite Hamiltonian $H=-\half\Delta$, with
$\Delta$ the Laplacian on $X$. 

Supersymmetry adds anticommuting variables, and for the supersymmetric
particle the Hilbert space is now the $L^2$-completion of the space of
differential forms on $X$,
\be
\cH = \W^*(X)
\ee
On this space we can realize the elementary $N=2$ supersymmetry
algebra
\be
[Q,Q^*]=-2H
\ee
by the use of the supercharge or differential $Q=d$ and its adjoint
$Q^*$. The spectrum of the Hamiltonian is encoded in the partition function
\be
Z(X;q,y) = \Tr_{\strut \cH} (-1)^F q^H y^F
\ee
with fermion number $F$ given by the degree of the differential form.

Of particular interest is the subspace $\V \subset \cH$ of
supersymmetric ground states, that satisfy $Q\psi=Q^*\psi=0$ and
therefore also $H\psi=0$. These zero-energy wavefunctions are
represented by harmonic differential forms
\be
\V = {\rm Harm}^*(X) \cong H^*(X).
\ee
We can compute the weighted number of ground states by the Witten
index, which defines a regularized superdimension\footnote{For a
graded vector space $V=V^+ \oplus V^-$ with even part $V^+$ and odd
part $V^-$, we define the superdimension as $\sdim V = \dim V^+ - \dim
V^-$, and, more generally, the supertrace of an operator $a$ acting on
$V$ as $\str_V(a)= \Tr_{V^+}(a) - \Tr_{V^-}(a)$. So we have $\sdim V =
\str_V1 = \Tr_V(-1)^F$. Here the Witten index operator $(-1)^F$
is defined as $+1$ on $V^+$ and $-1$ on $V^-$.} 
of the Hilbert space 
\be
\sdim \cH = \Tr_\cH (-1)^F =Z(X;q,1)
\ee
Since this expression does not depend on $q$, the Witten index simply
equals the Euler number of the space $X$
\be
\Tr_\cH (-1)^F  = \sdim \V = \sum_k (-1)^k \dim H^k(X) = \chi(X) .
\ee
Note that here we consider $H^*(X)$ as a graded vector space generated
by $b^+$ even generators and $b^-$ odd generators with
$\chi(X)=b^+-b^-$.

\newsubsection{Second-quantization and symmetric products}

The usual step of second-quantization now consists of considering a
system of $N$ of these (super)particles. It is implemented by the taking
the $N$-th symmetric product of the single particle Hilbert
space $\cH$
\be
S^N\cH = \cH^{\otimes n}/S_N,
\ee
or more properly the direct sum over all $N$
\be
\cQ_2 : \cH \ra S\cH = \bigoplus_{N\geq 0} S^N\cH.
\ee

We now propose to reverse roles. Instead of taking the symmetric
product of the Hilbert space of functions or differential forms on the
manifold $X$ (\ie, the symmetrization of the quantized manifold), we
will take the Hilbert space of functions or differential forms on the
symmetric product $\SNX$ (\ie, the quantization of the symmetrized
manifold)
\be
\cQ^2:\  X \ra SX=\coprod_{N\geq 0} \SNX.
\ee
The precise physical interpretation of this role-reversing is the
topic of these notes. It will appear later as a natural framework for
the light-cone quantization of string theory and of a certain class of
quantum field theories that are obtained as low-energy limits of
string theories. We will be particularly interested to learn whether these
operations commute (they will not)
\be
\commutingdiagram{X}{\cQ^1}{\cH}{\cQ^2}{\cQ^2}{SX}{\cQ^1}{S\cH}
\ee

But first we have to address the issue that the symmetric space $\SNX$
is not a smooth manifold but an orbifold, namely the quotient by the
symmetric group $S_N$ on $N$ elements,
\be
\SNX = X^N\!/S_N.
\ee
We will first be interested in computing the ground states for this
symmetric product, which we have seen are in general counted by the Euler number.
Actually, the relevant concept will turn out to be the orbifold Euler
number.  Using this concept there is a beautiful formula that was
first discovered by G\"ottsche \cite{goettsche} (see also \cite{gs,chea})
in the context of
Hilbert schemes of algebraic surfaces, but which is much more
generally valid in the context of orbifolds, as was pointed out by
Hirzebruch and H\"ofer \cite{hirzebruch}.

First some notation. It is well-known that many formulas for symmetric
products take a much more manageable form if we introduce generating
functions. For a general graded vector space we will use the notation
\be
S_p V = \bigoplus_{N\geq 0} p^N S^NV
\ee
for the weighted formal sum of symmetric products.
Note that for graded vector spaces the
symmetrization under the action of the symmetric group $S_N$ is
always to be understood in the graded sense, \ie, antisymmetrization for the
odd-graded pieces.
We recall that for an even vector space
\be
\dim S_pV = \sum_{N \geq 0} p^N \dim S^N V = (1-p)^{-\dim V}
\ee
whereas for an odd vector space
\be
\sdim S_pV = \sum_{N \geq 0} (-1)^N p^N \dim \ext^N V = (1-p)^{\dim V}
\ee
These two formulas can be combined into the single formula valid for
an arbitrary graded vector space that we will use often\footnote{This
can of course be generalized to traces of operators as $\str(S_p A)=
\sdet (1-p A)^{-1}$.}
\be
\sdim S_pV = (1-p)^{-\sdim V}
\ee
Similar we introduce for a general space $X$ the `vertex operator'
\be
S_pX = \hbox{`$\exp pX$'} = \coprod_{N\geq 0} \; p^N \SNX.
\ee
Using this formal expression the formula we are interested in reads
(see also \cite{vafa-witten})

\medskip

{\bf Theorem 1} \cite{goettsche,hirzebruch} --- {\sl
The orbifold Euler number
of the symmetric products $\SNX$ are given by}
\be
\chi_{orb}(S_p X)= \prod_{n>0} (1-p^n)^{-\chi(X)}.
\label{euler}
\ee

\newsubsection{The orbifold Euler character}

The crucial ingredient in Theorem 1 is the orbifold Euler character, a
concept that is very nicely explained in \cite{hirzebruch}. Here we
give a brief summary of its definition.

Suppose a finite group $G$ acts on a manifold $M$.  In general this
action will not be free and the space $M/G$ is not a smooth manifold
but an orbifold instead.  The {\it topological} Euler number of this
singular space, defined as for any topological space,
can be computed as the alternating sum of the
dimensions of the invariant piece of the cohomology,
\be
\chi_{top}(M/G) = \sdim H^*(M)^G.
\ee
In the de Rahm cohomology one can also simply take the complex of
differential forms that are invariant under the $G$-action and
compute the cohomology of the standard differential $d$.
This expression can be computed by averaging  over the group
\ba
\chi_{top}(M/G) \is  {1\over |G|} \sum_{g\in G}\; \str_{\strut \! H^*(M)} g \nonu
\is  {1\over |G|} \sum_{g\in G} \boxx g\id
\label{a}
\ea
Alternatively, using the Lefshetz fixed point formula, we can rewrite
this Euler number as a sum of fixed point contributions.  Let $M^g$
denote the fixed point set of the element $g\in G$. (Note that for the
identity $M^\id=M$.) Then we have
\ba
\chi_{top}(M/G) \is {1\over |G|}\sum_{g\in G} \; \sdim H^*(M^g) \nonu
\is  {1\over |G|} \sum_{g\in G} \boxx \id g
\label{b}
\ea
In the above two formulas we used the familiar string theory notation
\be
\boxx hg \; = \; \str_{\strut \! H^*(M^g)} h
\ee
for the trace of the group element $h$ in the `twisted sector' labeled
by $g$.  Note that the two expressions (\ref{a}) and (\ref{b}) for the
topological Euler number are related by a `modular
$S$-transformation,' that acts as
\be
\boxx g1 \ \ra\ \boxx 1g
\ee

The {\it orbifold} Euler number is the proper equivariant notion. We
see in a moment how it naturally appears in string theory. In the
orbifold definition we remember that on each fixed point set $M^g$
there is still an action of the centralizer or stabilizer subgroup
$C_g$ that consists of all elements $h \in G$ that commute with $g$.
The orbifold cohomology is defined by including the fixed point loci
$M^g$, but now taking only the contributions of the $C_g$
invariants. That is, we have a sum over the conjugacy classes $[g]$ of
$G$ of the topological Euler character of these strata
\be
\chi_{orb}(M/G) =  \sum_{[g]} \chi_{top}(M^g/C_g).
\ee
Note that this definition always gives an integer, in contrast with
other natural definitions of the Euler number of orbifolds.
From this point of view the topological Euler number only takes into
account the trivial class $g=\id$ (the `untwisted sector').  If we use
the elementary fact that $|[g]| = |G|/|C_g|$, we obtain in this way
\ba
\chi_{orb}(M/G) \is {1\over |G|} \sum_{g\in G}\;  \sdim H^*(M^g)^{C_g} \nonu
\is  {1\over |G|} \sum_{g,h \in G,\ gh=hg} \!\! \str_{\strut \! H^*(M^g)} 
h \nonu
\is  {1\over |G|} \sum_{g,h \in G,\ gh=hg} \!\!\boxx hg
\ea
This definition is manifest invariant under the `S-duality' that
exchanges $h$ and $g$.
We see that, compared with the topological definition, the orbifold
Euler number contains extra contribution of the `twisted sectors'
corresponding to the non-trivial fixed point loci $M^g$. Using again
Lefshetz's formula, it can be written alternatively in terms of the
cohomology of the subspaces $M^{g,h}$ that are left fixed by both $g$
and $h$ as
\be
\chi_{orb}(M/G) = {1\over |G|} \sum_{g,h \in G,\ gh=hg} \!\! \sdim
H^*(M^{g,h}).
\ee
It has been pointed out by Segal \cite{segal} that much of this and in
particular the applications to symmetric products that we are about to
give, find a natural place in equivariant K-theory. Indeed the
equivariant K-group (tensored with $\C$) of a space $M$ with a $G$
action is isomorphic to
\be
K_G(M) = \bigoplus_{[g]} K(M^g)^{C_g}.
\ee

\newsubsection{The orbifold Euler number of a symmetric product}

We now apply the above formalism to the case of the quotient
$X^N\!/S_N$.  For the topological Euler number the result is
elementary. We simply replace $H^*(X)$ by its symmetric product
$S^N\!H^*(X)$. Since we take the sum over all symmetric products,
graded by $N$, this is just the free symmetric algebra on the
generators of $H^*(X)$, so that \cite{macdonald}
\be
\chi_{top}(S_pX) =\sdim S_pH^*(X) = (1-p)^{-\chi(X)}
\ee

In order to prove the orbifold formula (\ref{euler}) we need to
include the contributions of the fixed point sets. Thereto we recall
some elementary facts about the symmetric group. First, the conjugacy
classes $[g]$ of $S^N$ are labeled by partitions $\{ N_n\}$ of $N$,
since any group element can be written as a product of elementary
cycles $(n)$ of length $n$,
\be
[g]=(1)^{N_1}(2)^{N_2}\ldots (k)^{N_k} ,\qquad \sum_{n>0} nN_n = N.
\ee
The fixed point set of such an element $g$ is easy to describe. The
symmetric group acts on $N$-tuples $(x_1,\ldots,x_N)\in X^N$.  A cycle
of length $n$ only leaves a point in $X^N$ invariant if the $n$ points
on which it acts coincide. So the fixed point locus of a general $g$
in the above conjugacy class is isomorphic to
\be
(X^N)^g \cong \prod_{n>0} X^{N_n}.
\ee
The centralizer of such an element is a semidirect product of 
factors  $S_{N_n}$ and $\Z_n$,
\be
C_g = S_{N_1} \times (S_{N_2} \ltimes \Z^{N_2}_2) \times \ldots 
(S_{N_k} \ltimes \Z^{N_k}_k). 
\ee
Here the factors $S_{N_n}$ permute the $N_n$ cycles $(n)$, while 
the factors $\Z_n$ act within one particular cycle $(n)$. The
action of the centralizer $C_g$ on
the fixed point set $(X^N)^g$ is obvious: only the subfactors $S_{N_n}$ act
non-trivially giving
\be
(X^N)^g/C_g \cong \prod_{n>0} S^{N_n} X.
\ee
We now only have to assemble the various components to compute the
orbifold Euler number of $\SNX$:
\ba
\chi_{orb}(S_pX) \is \sum_{N\geq 0} p^N \chi_{orb}(\SNX) \nonu
\is \sum_{N\geq 0} \, p^N \!\!\! \sum_{\scriptstyle \{N_n\} \atop
\scriptstyle \sum n N_n = N} \; \prod_{n>0} \;
\chi_{top}(S^{N_n}X) \nonu
\is \prod_{n>0}\; \sum_{N\geq 0} p^{nN} \chi_{top}(S^NX)\nonu
\is \prod_{n>0} \; (1-p^n)^{-\chi(X)}
\ea
which concludes the proof of ({\ref{euler}).

\newsubsection{Orbifold quantum mechanics on symmetric products}

The above manipulation can be extended beyond the computation of the
Euler number to the actual cohomology groups. We will only be able to fully
justify these definitions (because that's what it is at this point)
from the string theory considerations that we present in the next
section. For the moment let us just state that in particular cases where
the symmetric product allows for a natural smooth resolution (as for
the algebraic surfaces studied in \cite{goettsche} where the Hilbert
scheme provides such a resolution), we expect the orbifold definition
to be compatible with the usual definition in terms of the smooth
resolution.

We easily define a second-quantized, infinite-dimensional graded Fock
space whose graded superdimensions equal the Euler numbers that we
just computed. Starting with the single particle ground state Hilbert
space 
\be
\V= H^*(X)
\ee
we define it as the symmetric algebra of an
infinite number of copies $\V\nn$ graded by $n=1,2,\ldots$
\be
\cF_p = \bigotimes_{n>0} S_{p^n} \V\nn = 
S\Bigl(\bigoplus_{n>0} p^n \V\nn \Bigr).
\label{fock}
\ee
Here $\V\nn $ is a copy of $\V$ where the `number operator' $N$ is
defined to have eigenvalue $n$, so that
\be
\chi_{orb}(S_pX) = \Tr_{\strut \! \cF}(-1)^F p^N =  
\prod_{n>0} \; (1-p^n)^{-\chi(X)}.
\ee
We will see later that the degrees in $\V\nn $ are naturally shifted
by $(n-1){d\over 2}$ with $d$ the dimension of $X$, so that
\be
\V\nn  \cong H^{*-(n-1){d\over 2}}(X),\qquad n > 0.
\ee
Of course, this definition makes only good sense for even $d$, which
will be the case since we will always consider K\"ahler
manifolds. 

This result can be interpreted as follows. We have seen
that the fixed point loci consist of copies of $X$.  These copies
$X\nn $ appear as the big diagonal inside $S^nX$ where all $n$ points
come together. If we think in terms of middle dimensional cohomology,
which is particularly relevant for K\"ahler and hyperk\"ahler
manifolds, this result tells us that the middle dimensional cohomology
of $X$ contributes through $X^{(n)}$
to the middle dimensional cohomology of $S^nX$.

So, if we define the Poincar\'e polynomial as
\be
P(X;y) = Z(X;0,y)= \Tr_{\strut \V} (-1)^F y^F =
\sum_{0\leq k\leq d} (-1)^k y^{k}
b_k(X),
\ee
then we claim that the orbifold Poincar\'e polynomials of the
symmetric products $\SNX$ are given by
\be
P_{orb}(S_p X;y) =
\prod\twosub{n>0}{0\leq k\leq d}
\left(1-y^{k+ (n-1){d\over 2}} p^n\right)^{-(-1)^k b_k}
\ee
This is actually proved for the Hilbert scheme of an algebraic surface
in \cite{gs,chea}.

Although we will only be in a position to understand this well in the
next section, we can also determine the full partition function that
encodes the quantum mechanics on $\SNX$. Again the Hilbert space is a
Fock space built on an infinite number of copies of the single
particle Hilbert space $\cH(X)$
\be
\cH_{orb}(S_pX) = \bigotimes_{n>0} S_{p^n} \cH\nn (X).
\ee
The contribution to the total Hamiltonian of the states in the sector
$\cH\nn $ turns out to be scaled by a factor of $n$ relative to the
first-quantized particle, whereas the fermion number are shifted as
before, so that
\be
\cH\nn  \cong \W^{*-(n-1){d\over 2}}(X),\qquad
H\nn  = - \half \Delta/n.
\ee
To be complete explicit, let $\{h(m,k)\}_{m\geq 0}$ be the spectrum of
$H$ on the subspace $\W^k(X)$ of $k$-forms with
degeneracies\footnote{These degeneracies are consistently defined as
superdimensions of the eigenspaces, so that $c(m,k)\leq 0$ for $k$ odd, and
$c(0,k)=(-1)^kb_k$.}  $c(m,k)$, so that the single particle partition
function reads
\be
Z(X;q,y)= \Tr_{\strut \! \cH} (-1)^F y^F q^H = 
\sum\twosub{n>0}{0\leq k\leq d}
c(m,k) y^k q^{h(m,k)}.
\ee
Then we have for the symmetric product (in the orbifold sense)
\ba
Z_{orb}(S_p X;q,y) \is \Tr_{\strut \! \cH_{orb}(SX)} 
(-1)^F p^\cN q^H y^F \nonu
\is  \prod\twosub{n>0,\, m \geq 0}{0\leq k\leq d}\left(1-p^nq^{h(m,k)/n}
y^{k + (n-1){d\over 2}}\right)^{-c(m,k)}
\ea
In later sections we give a QFT interpretation of this formula.

\newsection{Second-quantized Strings}

The previous section should be considered just as a warming-up for the
much more interesting case of string theory. We will now follow all of
the previous steps again, going from a single quantized string to a
gas of second-quantized strings. In many respects this construction
--- in particular the up to now rather mysterious orbifold
prescription --- is more `canonical,' and all of the previous results
can be obtained as a natural limiting case of the string computations.

\newsubsection{The two-dimensional supersymmetric sigma model}

In the Lagrangian formulation the supersymmetric sigma model that
describes the propagation of a first-quantized string on a Riemannian
target space $X$ is formulated in terms of maps $x:\Sigma \ra X$ with
$\Sigma$ a Riemann surface, that we will often choose to give the
topology of a cylinder $S^1\times \R$ or a torus $T^2$. The canonical
Euclidean action, including the standard topological term, is of the
form
\be
{1\over 4\pi \a'} \int_\Sigma  G_{\m\n}(x) dx^\m \wedge *dx^\n +
{i\over 2\pi} \int_\Sigma x^*B + 
\hbox{\it fermions}
\ee
with $G$ the Riemannian metric and $B$ a closed two-form on $X$.

An important feature of the two-dimensional sigma model that in the
limit $\a'\ra 0$ it reduces to the supersymmetric quantum mechanics of
the previous section.  This limit can be equivalently seen as a
rescaling of the metric $G$ and thereby a low-energy or a large-volume
limit, $vol(X)\ra \infty$. In this point-particle limit the dependence
on the $B$-field disappears.

In the Hamiltonian formulation one describes a single string moving on
a space $X$ in terms of the loop space $\cL X$ of maps $S^1 \ra
X$. Depending on the particular type of string theory that we are
interested in, this first-quantization leads us to assign to the
manifold $X$ a single string SCFT Hilbert space
\be
\cQ^1 : X \ra \cH(X),
\ee
that can be formally considered to be the space of half-infinite
dimensional differential forms on $\cL X$.  We will always choose in
the definition of $\cH$ Ramond or periodic boundary conditions for the
fermions. These boundary conditions respect the supersymmetry algebra;
other boundary conditions can be obtained by spectral flow
\cite{spectralflow}.

On this Hilbert space act two natural operators: the Hamiltonian $H$,
roughly the generalized Laplacian on $\cL X$, and the momentum
operator $P$ that generates the canonical circle action on the loop
space corresponding to rotations of the loop,
\be
e^{i\th P}:\ x(\s) \mapsto x(\s + \th).
\ee
In a conformal field theory the operators $H$ and $P$ are usually
written in terms of left-moving and right-moving Virasoro
generators $L_0$ and $\Lbar_0$ as
\be
H = L_0+\Lbar_0-d/4, \qquad P = L_0 -\Lbar_0.
\ee
Here $d$ is the {\it complex} dimension of $X$.  If the manifold $X$
is a Calabi-Yau space, the quantum field theory carries an $N=2$
superconformal algebra with a $U(1)_L \times U(1)_R$ R-symmetry.  In
particular this allows us to define separate left-moving and
right-moving conserved fermion numbers $F_L$ and $F_R$, that up to an
infinite shift (that is naturally regularized in the QFT) 
represent the bidegrees
in terms of the Dolbeault differential forms on
$\W^*(\cL X)$.

The most general partition function is written as
\be
Z(X;q,y,\qbar,\ybar) = \Tr_{\strut \! \cH} (-1)^F   
 y^{F_L} \ybar^{F_R} q^{L_0-{d\over 8}}
\qbar^{\Lbar_0-{d\over8}}
\ee
with $F=F_L+ F_R$ the total fermion number.  
The partition function $Z$ represents the value of
the path-integral on a torus or elliptic curve, and we can write
$q=e^{2\pi i \tau}, y = e^{2\pi i z}$ with $\tau$ the modulus of the
elliptic curve and $z$ a point in its Jacobian that determines the
line-bundle of which the fermions are sections.  The spectrum of all
four operators $L_0,\Lbar_0,F_L,F_R$ is discrete with the further
conditions
\be
L_0,\Lbar_0 \geq {d/ 8},
\qquad L_0 - \Lbar_0 \in \Z,\qquad F_L,F_R \in \Z + \textstyle {d\over 2}.
\ee
For a general Calabi-Yau manifold it is very difficult to compute the
above partition function explicitly. Basically, only exact
computations have been done for orbifolds and the so-called Gepner
points, which are spaces with exceptional large quantum automorphism
groups.  This is not surprising, since even in the $\a'\ra 0$ limit we
would need to know the spectrum of the Laplacian, while for many
Calabi-Yau spaces such as $K3$ an explicit Ricci flat metric is not
even known.

Just as for the quantum mechanics case, we learn a lot by considering
the supersymmetric ground states $\psi \in V \subset \cH$ that satisfy
$H\psi=0$.  In the Ramond sector the ground states are canonically in
one-to-one correspondence with the cohomology classes in the Dolbeault
groups,
\be
\V \cong  H^{*,*}(X).
\ee
In fact, these states have special values for the conserved
charges. Ramond ground states always have $L_0=\Lbar_0 = d/8,$ and for
a ground state that corresponds to a cohomology class $\psi \in
H^{r,s}(X)$ the fermion numbers are shifted degrees
\be
F_L = r-d/2,\qquad F_R = s - d/2.
\label{fermion}
\ee
The shift in degrees by $d/2$ is a result from the fact that we had to
`fill up' the infinite Fermi sea. We see that there is an obvious
reflection symmetry $F_{L,R} \ra - F_{L,R}$ (Poincar\'e duality)
around the middle dimensional cohomology. If we take the limit
$q,\qbar \ra 0$, the partition function reduces essentially to the
Poincar\'e-Hodge polynomial of $X$
\ba
  h(X;y,\ybar) \is \lim_{q,\qbar \ra 0} Z(X; q,\qbar, y,\ybar) \nonu
\is \sum_{0\leq r,s\leq d} (-1)^{r+s}  y^{r-{d\over 2}} \ybar^{s-{d\over 2}} 
h^{r,s}(X)
\label{hodge}
\ea

\newsubsection{The elliptic genus}

An interesting specialization of the sigma model partition function is
the elliptic genus of $X$ \cite{ell}, defined as
\be
\chi(X;q,y) = \Tr_{\strut \! \cH} 
(-1)^F  y^{F_L} q^{L_0 - {d\over 8}}
\ee
The elliptic genus is obtained as a specialization of the general
partition function for $\ybar=1$. Its proper definition is
\be
\chi(X;q,y) = \Tr_{\strut \! \cH} 
(-1)^Fy^{F_L} q^{L_0 - {d\over 8}}\qbar^{\Lbar_0 - {d\over 8}} 
\ee
But, just as for the Witten index, because of the factor $(-1)^{F_R}$
there are no contributions of states with $\Lbar_0 - d/8>0$. Only the
right-moving Ramond ground states contribute. The genus is therefore
holomorphic in $q$ or $\t$. Since this fixes $L_0 - d/8$ to be an
integer, the partition function becomes a topological index, with no
dependence on the moduli of $X$.

Using general facts of modular invariance of conformal field theories,
one deduces that for a Calabi-Yau $d$-fold the elliptic genus is a
weak Jacobi form \cite{eichler} of weight zero and index $d/2$. (For
odd $d$ one has to include multipliers or work with certain finite
index subgroups, see \cite{neumann,kawai,kawai-2}.) The ring of Jacobi forms
is finitely generated, and thus finite-dimensional for fixed
index\footnote{For example, in the case $d=2$ it is one-dimensional
and generated by the elliptic genus of $K3$.}.  It has a Fourier
expansion of the form
\be
\chi(X;q,y) = \sum_{m\geq 0,\;\ell} c(m,\ell) q^my^\ell
\ee
with integer coefficients. The terminology `weak' refers to the fact
that the term $m=0$ is included. 

The elliptic genus has beautiful mathematical properties.  In contrast
with the full partition function, it does not depend on the moduli of
the manifold $X$: it is a (differential) topological invariant.  In
fact, it is a genus in the sense of Hirzebruch --- a ring-homomorphism
from the complex cobordism ring $\W^*_U(pt)$ into the ring of weak
Jacobi forms. That is, it satisfies the relations
\ba
\chi(X\cup X';q,y) \is \chi(X;q,y) + \chi(X';q,y), \nonu 
\chi(X\times X';q,y) \is \chi(X;q,y)\cdot \chi(X';q,y),\\[2mm]
\chi(X;q,y) \is 0,\qquad\mbox{if $X=\partial Y$,} \nonumber
\ea
where the last relation is in the sense of complex bordism. The first
two relations are obvious from the quantum field theory point of view;
they are valid for all partition functions of sigma models.  The last
condition follows basically from the definition in terms of classical
differential topology, more precisely in terms of Chern classes of
symmetrized products of the tangent bundle, that we will give in a
moment.  We already noted that in the limit $q \ra 0$ the genus
reduces to a weighted sum over the Hodge numbers, which is essentially
the Hirzebruch $\chi_y$-genus,
\be
\chi(X;0,y) =\sum_{r,s} (-1)^{r+s} h^{r,s}(X)y^{r-{d\over 2}},
\ee
and for $y=1$ its equals the Witten index or Euler number of $X$
\be
\chi(X;q,1)=\Tr_\cH (-1)^F = \chi(X).
\ee
For smooth manifolds, the elliptic genus  has an equivalent definition as
\be
\chi(X;q,y)=\int_X  ch(E_{q,y}) td(X)
\ee
with the formal sum of vector bundles
\be
E_{q,y} =  y^{-{d\over 2}} \bigotimes_{n > 0}\Bigl( \ext_{-yq^{n-
1}}T_X\otimes \ext_{-y^{-1}q^n} \overline{T}_X \otimes S_{q^n} T_X\otimes 
S_{q^n} \overline{T}_X\Bigr),
\ee
where $T_X$ denotes the holomorphic tangent bundle of $X$. If the
bundle $E_{y,q}$ is expanded as
\be
E_{q,y} = \bigoplus_{m,\;\ell} q^m y^\ell E_{m,\ell}
\ee
the coefficients $c(m,\ell)$ give the index of the Dirac operator on
$X$ twisted with the vector bundle $E_{m,\ell}$, and are therefore
integers. This definition follows from the sigma model by taking the
large volume limit, where curvature terms can be ignored and one
essentially reduces to the free model, apart from the zero modes that
give the integral over $X$.

\newsubsection{Physical interpretation of the elliptic genus}

Physically, the elliptic genus arises in two interesting
circumstances. First, it appears as a counting function of
perturbative string BPS states. If one constrains the states of the
string to be in a right-moving ground state, \ie, to satisfy
$\Lbar_0 = d/8$, the states are invariant under part of the space-time
supersymmetry algebra and called BPS. The generating fucntion of such
states is naturally given by the elliptic genus. Because we weight the
right-movers with the chiral Witten index $(-1)^{F_R}$, only the
right-moving ground states contribute.

Another physical realization is the so-called half-twisted model.
Starting from a $N=2$ superconformal sigma model, we can obtain a
topological sigma model, by changing the spins of the fermionic
fields.  This produces two scalar nilpotent BRST operators $Q_L, Q_R$
that can be used to define cohomological field theories. If we use
both operators, or equivalently the combination $Q=Q_L+Q_R$, the
resulting field theory just computes the quantum cohomology of
$X$. This topological string theory is the appropriate framework to
understand the Gromov-Witten invariants. If we ignore interactions for
the moment, the free spectrum is actually that of a quantum field
theory. Indeed, the gauging implemented by the BRST operator removes
all string oscillations, forcing the states to be both left-moving
and right-moving ground states
\be
L_0\psi=\Lbar_0\psi=0.
\ee
Only the harmonic zero-modes contribute. In this way one finds one
quantum field for every differential form on the space-time. This is
precisely the model that we discussed in the previous section. 

However, as first suggested by Witten in \cite{witten-mirror}, it is
also possible to do this twist only for the right-moving fields. In
that case, we have to compute the cohomology of the right-moving BRST
operator $Q_R$. This cohomology has again harmonic representatives
with $\Lbar_0=0$. These states coincide with the BPS states mentioned
above. The half-twisted cohomology is no longer finite-dimensional,
but it is graded by $L_0$ and $F_L$ and the dimensions of these graded
pieces are encoded in the elliptic genus.  So, the half-twisted string
is a proper string theory with an infinite tower of heavy states.

\newsubsection{Second-quantized elliptic genera}

We now come to analogue of the theorem of G\"ottsche and Hirzebruch for
the elliptic genus as it was conjectured in \cite{dyon} and derived in 
\cite{ell-genus}. 

\medskip

{\bf Theorem 2} \cite{ell-genus} --- {\sl
The orbifold elliptic genus of the symmetric
products $\SNX$ are given by}
\be
\chi_{orb}(S_pX;q,y) = 
\prod_{n>0,\,m\geq 0,\,\ell} (1-p^nq^m y^\ell)^{-c(nm,\ell)}
\label{genus}
\ee

\medskip

In order to prove this result, we have to compute the elliptic genus
or, more generally, the string partition function for the orbifold
$M/G$ with $M=\SNX$ and $G=S_N$. The computation follows closely the
computation of the orbifold Euler character that was relevant for the
point-particle case.

First of all, the decomposition of the Hilbert space in superselection
sectors labeled by the conjugacy class of an element $g\in G$
follows naturally. The superconformal sigma model with target space
$M$ can be considered as a quantization of the loop space $\cL
M$. If we choose as our target space an orbifold $M/G$,
the loop space $\cL(M/G)$ will have disconnected components of loops in
$M$ satisfying the twisted boundary condition
\be
x(\s+2\pi)=g\. x(\s),\qquad g \in G,
\ee
and these components are labeled by the conjugacy classes $[g]$.  In
this way, we find that the Hilbert space of any orbifold
conformal field theory decomposes naturally into twisted sectors.
Furthermore, in the untwisted sector we have to take the states that are
invariant under $G$. For the twisted sectors we can only take invariance
under the centralizer $C_g$, which is the largest subgroup that commutes
with $g$. If $\cH_g$ indicates the sector twisted by $g$, the orbifold
Hilbert space has therefore the general form \cite{orbifold}
\be
\cH(M/G) = \bigoplus_{[g]} \cH_g^{C_g}
\ee
In the point-particle limit $\a'\ra 0$ the size of all loops shrinks
to zero. For the twisted boundary condition this means that the loop
gets necessarily concentrated on the fixed point set $M^g$ and we are
in fact dealing with a point-particle on $M^g/C_g$. In this way the
string computation automatically produces the prescription for the
orbifold cohomology that we discussed before. Indeed, as we stress,
the quantum mechanical model of the previous section can best be
viewed as a low-energy limit of the string theory.

In the case of the symmetric product $\SNX$, the orbifold
superselection sectors correspond to partitions $\{N_n\}$ of
$N$. Furthermore, we have seen that for a given partition the fixed
point locus is simply the product
\be
\prod_n S^{N_n}X\nn 
\ee
Here we introduce the notation $X\nn $, to indicate a copy of $X$
obtained as the diagonal in $X^N$ where $n$ points coincide. In the
case of point-particles this distinction was not very important but
for strings it is absolutely crucial.

The intuition is best conveyed with the aid of 
\figuurplus{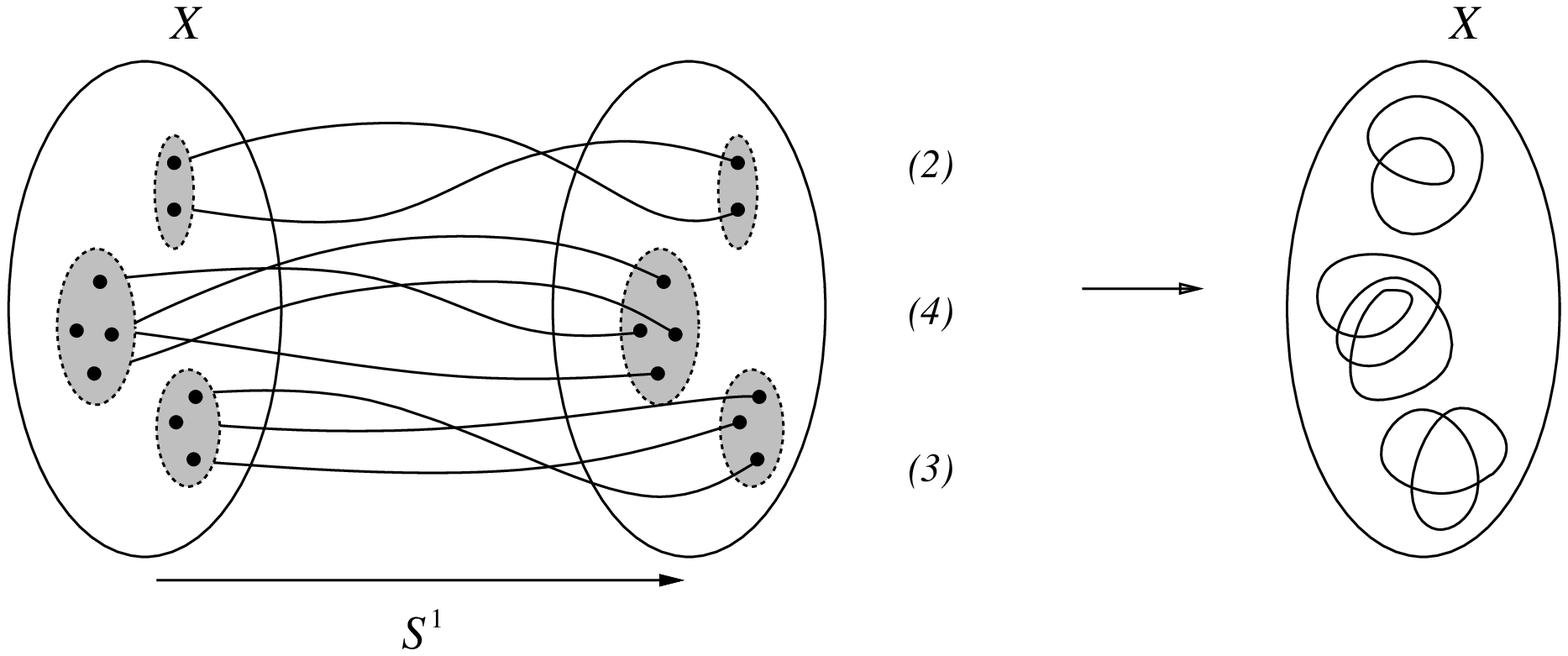}{12cm}{A twisted sector of a sigma model on $\SNX$ can
describe less than $N$ strings. (Here $N=9$ and the sector contains
three `long strings.')} where we depicted a generic
twisted sector of the orbifold sigma model. The crucial point is that
such a configuration can be interpreted as describing long
strings\footnote{The physical significance of this picture was developed
in among others \cite{strominger-vafa,maldacena-susskind,fivebrane}
and made precise in \cite{ell-genus}.}
whose number can be smaller than $N$. Indeed, as we clearly see, a
twisted boundary condition containing a elementary cycle of length $n$
gives rise to a single string of `length' $n$ built out of $n$ `string
bits.' If the cycle permutes the coordinates $(x_1,\ldots,x_n) \in X^n$
as
\be
x_k(\s+2\pi) = x_{k+1}(\s),\qquad k\in(1,\ldots,n),
\label{twist}
\ee
we can construct a new loop $x(\s)$ by gluing the $n$ strings
$x_1(\s),\ldots,x_n(\s)$ together:
\be
x(\s)=x_k(\s') \quad \hbox{if} \quad 
\textstyle \s={1\over n}\bigl(2\pi (k-1)+ \s'\bigr),\
\s'\in[0, 2\pi].
\ee
If the twist element is the cycle $(N) \in S_N$, such a configuration
describes one single long string of length $N$, instead of the $N$
short strings that we would expect.

In this fashion we obtain from a cyclic twist $(n)$ one single copy of
the loop space $\cL X$ that we denote as $\cL X\nn $.  We use the
notation $\cH\nn $ for its quantization.  The twisted loop space
$\cL X\nn $ is distinguished from the untwisted loop space $\cL X$
in that the canonical circle action is differently normalized. We now
have
\be
e^{i\th P}:\ x(\s) \ra x(\s + \th/n).
\ee
So we find that only for $\th=2\pi n$ do we have a full rotation
of the loop.
This is obvious from the twisted boundary condition (\ref{twist}).
It seems to imply that the eigenvalues for the operator 
$P=L_0-\Lbar_0$ in this
sector are quantized in units of $1/n$. Together with the
fact that in the elliptic genus only states with $\Lbar_0 = 0$ contribute,
this would suggest that
the contribution of the sector $\cH\nn $ to the elliptic genus 
is\footnote{We use the more general notation $\chi(\cH;q,y) = \Tr_{\! \cH} 
(-1)^F q^{L_0 - {d\over 8}} y^{F_L}$ for any Hilbert space $\cH$.} 
\be
\chi(\cH\nn ;q,y) \mathop{=}^? \sum_{m,\ell} c(m,\ell) q^{m/n}y^\ell.
\ee
However, we must remember that the centralizer of a cycle of length
$n$ contains a factor $\Z_n$. This last factor did not play a role in
the point-particle case, but here it does act non-trivially. In fact,
it is precisely generated by $e^{2\pi i P}$. The orbifold definition
includes a prescription to take the states that are invariant under
the action of the centralizer. So only the states with integer
eigenvalues of $P$ survive. In this way only the states with $m$
congruent to $0$ modulo $n$ survive and we obtain an integer
$q$-expansion,
\be
\chi(\cH\nn ;y,q) = \sum_{m,\ell} c(nm,\ell) q^{m}y^\ell.
\ee
We now again assemble the various components to finish the proof
of (\ref{genus}) (for more details see \cite{ell-genus}). 
\ba
\sum_{N\geq 0} p^N \chi_{orb}(\SNX;q,y) \is 
\sum_{N\geq 0} \, p^N \!\!\! \sum\twosub{N_n}{\sum n N_n = N} 
\prod_{n>0} \; \chi(S^{N_n}\cH\nn ;q,y) \nonu
\is \prod_{n>0}\; \sum_{N\geq 0} p^{nN}  \chi(S^N \cH\nn ;q,y)\nonu
\is \prod_{n>0,\;m,\;\ell} \; (1-p^nq^my^\ell)^{-c(nm,\ell)}
\ea
The infinite product formula has strong associations to automorphic
forms and denominator formulas of generalized Kac-Moody algebras
\cite{borcherds} and string one-loop amplitudes \cite{harvey-moore}; we
will return to this.

\newsubsection{General partition function}

It is not difficult to repeat the above manipulations in symmetric algebra
for the full partition function. In fact, we can write a general
formula for the second-quantized string Fock space, similar as
we did for the point-particle case in (\ref{fock}). This Fock space
is again of the form
\be
\cF_p = \bigotimes_{n>0} S_{p^n} \cH\nn .
\ee
Here $\cH\nn $ is the Hilbert space obtained by
quantizing a single string that is wound $n$ times.
It is isomorphic to the subspace of the single string Hilbert
space $\cH=\cH^{(1)}$ with 
\be
L_0-\Lbar_0 =0\ \hbox{(mod $n$).}
\label{modn}
\ee
The action of the operators $L_0$ and $\Lbar_0$ on
$\cH\nn $ are then rescaled by a factor $1/n$ compared with
the action on $\cH$
\be
L_0\nn  = {L_0^{(1)}/n},\qquad 
\Lbar_0\nn  = {\Lbar_0^{(1)}/n}.
\ee
As we explained already, this rescaling is due to the fact that the
string has now length $2\pi n$ instead of $2\pi$. Even though the
world-sheet Hamiltonians $L_0\nn  ,\Lbar_0\nn  $ have fractional
spectra compared to the single string Hamiltonians, the momentum
operator still has an integer spectrum,
\be
L_0\nn - \Lbar_0\nn =0\ \hbox{(mod 1),}
\ee
due to the restriction (\ref{modn}) that is implemented by the
orbifold $\Z_n$ projection.

It is interesting to reconsider the ground states of $\cH\nn $ in
particular their $U(1)_L\times U(1)_R$ charges, since this will teach
us something about the orbifold cohomology of $\SNX$. A 
ground state $\psi\nn  \in
\V\nn $ that correspond to a cohomology class 
$\psi\in H^{r,s}(X)$ still has fermion charges $F_L,F_R$ given by
\be
F_L = r-d/2,\qquad F_R = s - d/2.
\ee
Making the string longer does not affect the $U(1)$ current algebra.
However, since these states now appear as a ground states of a
conformal field theory with target space $S^n\! X$, which is of
complex dimension $n\.d$, these fermion numbers have a different
topological interpretation.  The corresponding degrees $r\nn,s\nn$ of
the same state now considered as a differential form in the orbifold
cohomology of $S^n\! X \subset \SNX$ are therefore shifted as
\be
r\nn = r + (n-1){d/ 2},\qquad
s\nn = s + (n-1){d/ 2}.
\ee
That is, we have
\be 
\V\nn \cong  H^{*-(n-1){d\over 2},*-(n-1){d\over 2}}(X).
\ee
In the quantum mechanics limit, the twisted loops that give rise to
the contribution $\cH\nn $ in the Fock space become point-like and
produce another copy $X\nn $ of the fixed point set $X$. However, this
copy of $X$ is the big diagonal in $X^n$. We see that this gives
another copy of $H^*(X)$ however now shifted in degree. In the full
Fock space we have an infinite number of copies, shifted by positive
multiples of $(d/2,d/2)$. 

We can encode this all in the generating function of Hodge numbers
(\ref{hodge}) as
\be
h_{orb}(S_pX;y,\ybar) = \prod_{n>0,\, r,\, s}
\left(1-p^ny^r\ybar^s\right)^{-(-1)^{r+s}h^{r,s}(X)}
\ee
For the full partition function we can write a similar expression
\be
Z_{orb}(S_pX;q,\qbar,y,\ybar) = \prod_{n>0} \!\! \prod\twosub{h,\, \bar{h},\, r,\, s}
{h-\bar h= 0\ ({\rm mod}\ n)}\!\!\!\!\!
\left(1-p^nq^h\qbar^{\bar h} y^r\ybar^s\right)^{- c(h,\bar h,r,s)}
\ee
where
\be
Z(X;q,\qbar,y,\ybar) = \sum_{h,\, \bar h,\, r,\, s}
c(h,\bar h,r,s)q^h\qbar^{\bar h} y^r \ybar^s
\ee
is the single string partition function.

\newsection{Light-Cone Quantization of Quantum Field Theories}

We now turn to the physical interpretation of the above results. Usually
quantum field theories are quantized by splitting, at least locally, a
Lorentzian space-time in the form $\R \times \Sigma$ where $\R$
represents the time direction and $\Sigma$ is a space-like Cauchy
manifold. Classically, one specifies initial data on $\Sigma$ which then
deterministically evolve through some set of differential equations in
time. In recent developments it has proven useful to use for the time
direction a null direction. This complicates of course the initial value
problem, but has some other advantages. One can try to see this a
limiting case where one uses Lorentz transformations to boost the
time-like direction to an almost null direction \cite{seiberg}.

\newsubsection{The two-dimensional free scalar field revisited}

Before we turn to the interpretation of our results on second
quantization and symmetric products in terms of quantum field theory
and quantum string theory, let us first revisit one of the simplest
examples of a QFT and compute the partition function of a
two-dimensional free scalar field. In fact, let us be slightly more
general and consider a finite number $c$ of such scalar fields labeled
by a $c$-dimensional real vector space $V$. (One could easily take
this vector space to be graded, but for simplicity we assume it to be
even.)

Quantization of this model usually proceeds as follows: one chooses a
two-dimensional space-time with the topology of a cylinder $\R \times
S^1$ and with coordinates $(x^0,x^1)$. One then introduces the light-cone
variables $x^\pm=x^0 \pm x^1$. A classical solution of the equation of
motion $\Delta \f = 0$ is decomposed as
\be
\f(x^+,x^-) = q + p x^0 + \f_L(x^-) + \f_R(x^+),
\ee
where the zero-mode contribution (describing a point particle on $V$
with coordinate $q$ and conjugate momentum $p$) is isolated from the
left-moving and right-moving oscillations $\f_L(x^-)$ and
$\f_R(x^+)$. The non-zero modes have a Fourier expansion
\be
\f_L(x^-) = \sum_{n \not=0 } {\textstyle {1\over n}}\a_n e^{inx^-}
\ee
with a similar expression for $\f_R(x^+)$.

In canonical quantization the Fourier oefficients $\a_n$ are replaced
by creation and annihilation operators with commutation relations
$[\a_n,\a_m]=n\delta_{n+m}$. This Heisenberg algebra is realized on a
Fock space $\cF$ built on a vacuum state $|0\>$ satisfying
$\a_n|0\>=0$ for $n>0$. This Fock space can be written in terms of
symmetric products as
\be
\cF_p = \bigotimes_{n>0} S_{p^n} V\nn 
=S^*\Bigl(\bigoplus_{n>0} p^n V\nn \Bigr),
\label{F}
\ee
where $V\nn $ is a copy of $V$ with the property that the (chiral)
oscillation number operator $N$ has eigenvalue $n$ on $V\nn $. 
The $p$-expansion keeps track of the $N$-gradation.
As a quantum operator $N$ is defined as
\be
N = \sum_{n>0} \a_{-n}\a_n.
\ee
The chiral partition function is now written as a character of this
module (with $p=e^{2\pi i \tau}$, $\t$ in the upper half-plane $\H$)
\be
\chi(p) = \dim \cF_p = \Tr_{\strut \!\cF} p^{N} = \prod_{n>0} (1-p^n)^{-c}
\ee
This character is almost a modular form of weight $-c/2$. One way to see
this, is by considering the partition function
of the full Hilbert space $\cH$, which is modular invariant. $\cH$ is
obtained by combining the left-moving
oscillators with the right-moving oscillators and adding the zero-mode
contribution
\be
\cH = L^2(V) \otimes \cF \otimes \bar\cF
\ee
The total chiral Hamiltonians can be written as
\be
L_0 = -\half\Delta + N ,\qquad
\Lbar_0 = -\half\Delta + \bar N.
\ee
The full partition function is then evaluated as
\be
Z(p,\pbar) = \Tr_{\strut \! \cH} p^{L_0-c/24} \pbar^{\Lbar_0-c/24} = 
\left(\sqrt{\Im \t}|\eta(p)|^2\right)^{-c}
\label{par}
\ee
with $\eta(p)$ the Weierstrass eta-function
\be
\eta(p)= p^{1\over 24} \prod_{n>0} (1-p^n).
\ee
So the proper modular object is given by
\be
\Tr_{\strut \cF}p^{N-c/24} = \eta(p)^{-c}
\ee
which is a modular form of $SL(2,\Z)$ of weight $-c/2$ (with multipliers
if $c \not= 0$ (mod 24).) The extra factor $p^{-c/24}$ is interpreted
as a regularized infinite sum of zero-point energies that appear in canonical
quantization.

In the Lagrangian formalism the same result is given in terms of a
$\zeta$-function regularized determinant of $c$ scalar fields
\be
Z = (\sqrt \Im \t/\det{}' \Delta)^{c/2} 
\ee
with $\Delta$ the laplacian on the torus $T^2$ and the prime indicates
omission of the zero-mode. This determinant can be computed in a
first-quantized form as a one-loop integral
\be
\log Z = -\half \log\det \Delta = - \half \Tr \log \Delta = 
\int_0^\infty  {dt\over t} \, \Tr_{\strut \!\cH} 
e^{-tH}
\ee
with $\cH=L^2(T^2)$ is now the quantum mechanical Hilbert space of
a single particle moving on $T^2$. Here
the RHS is defined by cutting of the integral at $t=\e$ and
subtracting the $\e$-dependent (but $\t$-independent) term.

Let us mention a few aspects of these results that we will try to
generalize when we consider strings instead of quantum fields in
the next section.

\begin{enumerate}
\item The quantum field theory partition function factorizes in left-moving
and right-moving contributions that are holomorphic functions of the
modulus $p$. 

\item The holomorphic contributions are modular forms of weight $-c/2$
under $SL(2,\Z)$ if a particular correction (here $p^{-c/24}$) is added.

\item The full partition fucntion is modular invariant because the
zero-mode contribution adds a non-holomorphic factor $(\Im\t)^{-c/2}$.

\item The holomorphic contributions are characters of an
infinite-dimensional Kac-Moody algebra, in fact in this simple case just
the Heisenberg algebra generated by the operators $\a_n$. 

\item The modularity of the characters, \ie, the transformation
properties under the modular group $SL(2,\Z)$ is `explained' by the
relation to a partition function of a quantum field on a two-torus $T^2$
with modulus $\tau$ and automorphism group $SL(2,\Z)$.

\end{enumerate}

\newsubsection{Discrete light-cone quantization}

In light-cone quantization one works on $\R^{1,1}$ in terms of the
light-cone coordinates $x^\pm$ with metric
\be
ds^2 = 2 dx^+ dx^-,
\ee
but now chooses the null direction $x^+$ as the time coordinate. We
will write the conjugate momenta as
\be
p^+ = p_- = -i{\d\over \d x^-},\qquad
p^- = p_+ = -i{\d\over \d x^+}.
\ee
In the usual euclidean formulation of two-dimensional CFT we have $p^+
= L_0,$ $p^-=\Lbar_0$.  In this setup a free particle has an eigentime
given by $x^+ = p^+ t$.  The light-cone Hamiltonian describes
evolution in the `light-cone time' $x^+$ and so is given by
\be
H_{lc}=p^-
\ee
An initial state is specified by the $x^-$-dependence for fixed $x^+$. 

The so-called discrete light-cone quantization (DLCQ) further assumes
that the null direction $x^-$ is compact of radius $R$
\be
x^- \sim x^- + 2\pi R.
\ee
(The specific value of $R$ is not very important since
it can of course be rescaled by a Lorentz
boost. We will therefore
often put it to one, $R=1$.) We denote the Lorentzian manifold so obtained as 
$(\R\times S^1)^{1,1}$.  The periodic identification of $x^-$
makes the spectrum of the conjugate
momentum $p^+$ discrete
\be
p^+ \in N/R,\qquad N\in \Z.
\ee
Now we have for fixed $x^+$ a decomposition of the scalar field as
\be
\f(x^-) = \sum_{n \not=0 } {\textstyle {1\over n}}\a_n e^{inx^-}
\ee
(Clearly, this quantization scheme is incomplete, since
we are omitting the zero-modes with $p^+=0$. We will only obtain the
left-moving sector of the theory. We will return to this point.)
Since the classical equation for a free field reads
\be
\d_+\d_- \f = 0,
\ee
the field $\f(x^-)$ will have no $x^+$-dependence and therefore
the light-cone energy of its modes will vanish, $p^-=0$. 
If we have $c$ of these free scalar fields $\f(x)\in V$, 
quantization will result in the same chiral Fock space 
that we considered in the canonical quantization
\be
\cF_p = \bigotimes_{n>0} S_{p^n}V^{(n)}
\ee
and the light-cone partition function is given by same
infinite product
\be
\Tr_{\strut \!\cF }\, p^{P^+} = \prod_{n>0} (1-p^n)^{-c}
\ee
Note that the eigenvalues of the longitudinal momentum $P^+=n$ are always
positive. This is due to the fact that the oscillation
numbers $\a_n$ form a Heisenberg algebra.

We recognize in these formulas our computations of the
Euler number of the symmetric products of a space $X$ with
$\chi(X)=c$. To explain this relation we now consider the
light-cone quantization of field theories in higher dimensions.

\newsubsection{Higher-dimensional scalar fields in DLCQ}

Things become a bit more interesting if we consider a free scalar
field on a more general space-time of the form
\be
M^{1,d+1} = (\R\times S^1)^{1,1}\times X^d,
\ee
$X$ compact Riemannian and with light-cone coordinates $(x^+,x^-,x)$.
We adopt light-cone quantization and consider as initial data the
field configuration on $x^+=\,${\it constant.} The light-cone
Hamiltonian $p^-$ again describes the evolution in $x^+$. It will be
convenient to perform a Fourier transformation in the light-cone
coordinate $x^-$ and consider a basis of field configurations of the form
\be
\f(x^+,x^-,x) = e^{i(p^-x^+ + p^+ x^-)} \f_m(x)
\ee
with $p^+=n$ (we put $R=1$) and $\f_m(x)$ an eigenstate of the transverse
Hamiltonian $H=-\half \Delta^{(X)}$, 
\be
H\phi_m = h_m \phi_m.
\ee
The equation of motion on the space-time $M$,
 $\Delta^{(M)}\f=0$, then gives the so-called mass-shell
relation
\be
p^- = - {1\over 2 p^+}\Delta^{(X)}= {1\over n} h_m .
\ee
Here we see an interesting phenomenon. The light-cone energy is given
by a non-relativistic expression of the form $p^2/2m$, where $p$ is
the transversal momentum  and the `mass' $m$ is given by the longitudinal
momentum $p^+=n$. (On a curved manifold $p^2$ is replaced by the
eigenvalues $h_m$ of the Laplacian.)
The appearance of this non-relativistic expression has it geometric
explanation in the fact that the stabilizer group of a null-direction in
$\R^{1,n+1}$ is the Galilean group of $\R^n$. The formula implies that
for a particle with $p^+ = n$, the light-cone energy $p^-$ is rescaled by
a factor of $1/n$.

Note that quite generally in light-cone quantization the symmetries
of the underlying space-time manifold are not all manifest. If we work with the
Minkowski space-time $\R^{1,n+1}$ the Lorentz group $SO(1,n+1)$ is
partly non-linearly realized.  For interacting QFT's the proof of
Lorentz invariance of a light-cone formulation is highly non-trivial.
In DLCQ the Lorentz-invariance is only expected in the limit $R\ra
\infty$. Since the value of $R$ can be rescaled by a Lorentz boost, this
limit is equivalent to the large $N$ limit, $N\ra \infty$. Again, for
interacting theories the appearance of Lorentz-invariance in this limit
is not obvious.

Upon quantization we obtain in the present case
a Fock space that is of the form
\be
\cF_p = \bigotimes_{n>0} \bigoplus_{N\geq 0} p^{nN} S^N \cH\nn 
\ee
with $\cH\nn=L^2(X)$ with $p^-=H\nn=H/n$ the rescaled QM Hamiltonian. 
Now the light-cone energies
$p^-$ will not typically vanish, and the full partition function
is given by a two-variable function,
\be
Z(X,p,q)= \Tr_{\strut \cF}p^{P^+}q^{P^-},
\ee
with $P^+,P^-$ the total light-cone momentum operators (with
eigenvalues $p^+,p^-$). 

From the above description is should have become clear that this
partition function can be completely identified with the quantum
mechanics on the symmetric product space $S_pX= \coprod_N p^N\SNX$ that we
discussed in such details in section 2. We therefore obtain:

\medskip

{\bf Theorem 3} --- {\sl The discrete light-cone quantization of a free 
scalar field
on the space-time $M=(\R\times S^1)^{1,1}\times X$ with total
longitudinal momentum $p^+=N$ is 
given by the quantum mechanics on the orbifold symmetric product $\SNX$,
\be
\cH^{QFT}(X) = \cH^{QM}_{orb}(SX).
\ee
Furthermore, the light-cone Hamiltonian $p^-$ is identified with the
non-relativistic quantum mechanics Hamiltonian $H$.}

\newsubsection{The supersymmetric generalization}

It is easy to extend this construction to a physical system that
describes the supersymmetric quantum mechanics on $SX$. In that case we
want to have arbitrary differential forms on $X$, so our fundamental
fields will be free $k$-forms $\f^k \in \W^k(M)$ with $0\leq k \leq d$
on the space-time $M=(\R\times S^1)^{1,1}\times X$. These fields
have a quadratic
action (with fermionic statistics if $k$ is odd)
\be
\int _Y \half d\phi \wedge * d\phi,\qquad \f=\sum_k \f^k\in\W^*(Y).
\ee
This gives as equation of motion the Maxwell equation
\be
d^*d\f = 0.
\ee
This Lagrangian is invariant under the gauge symmetry
\be
\phi^k \ra \phi^k + d \l^{k-1},\qquad \l \in \W^{k-1}(M),
\label{gauge}
\ee
giving $\f^k$ the interpretation of a generalized $k$-form
connection with curvature $d\f^k$. 
This gauge symmetry can be fixed by requiring
\be
\iota_{\d/\d x^+}\f =0,
\ee
a condition that we write as $\f_+=0$. With this gauge
condition the equation of motion can be used to eliminate
the component $\f_-$ in terms of the transversal components
$\f\in\W^*(X)$, leaving only the form on the transverse space 
$X$ as physical. All of this is well-known from the description of
the RR fields of the light-cone type II superstring (or supergravity).

With this gauge fixing we naturally reduce the second-quantized
light-cone description to supersymmetric quantum mechanics on $SX$. We
therefore find exactly the field theoretic description of our SQM model
of section 2. It describes the multi-form abelian gauge field theory on
$M=(\R\times S^1)^{1,1}\times X$ in DLCQ, in particular we have
\be
\cH^{QFT}(X) = \cH^{SQM}_{orb}(S_pX)=\bigotimes_n S_{p^n}\W^*(X)\nn
\ee
where powers of $p$ keep track of the longitudinal momentum $p^+$.

Particular interesting is the zero-energy $p^-=0$ sector $\V^{QFT}
\subset \cH^{QFT}$.
Since $p^-$ is identified with the SQM Hamiltonian,
these states correspond the ground states of the supersymmetric
quantum mechanics and we have
\be
\V^{QFT} = \bigotimes_{n>0} S_{p^n} H^*(X)
\ee
and the partition function of this zero $p^-$ sector
reproduces exactly the orbifold Euler character
\be
\Tr_{\strut \V_{QFT}}(-1)^F p^{P^+} = \chi_{orb}(S_pX).
\ee
The modular properties are now explained along the lines of section 1.
This partition function can be computed in a Lagrangian formulation
by considering the compact space-time $T^2 \times X$. The explicit
$T^2$ factor explain the occurrence of $SL(2,\Z)$.

The modular properties are particularly nice
if we choose as our manifold $X$ to be a $K3$ surface with $\chi(X)=24$.  We
then almost have a modular object without multipliers,
\be
\chi_{orb}(S_pX) = {p\over \Delta(p)}
\ee
with $\Delta(p)= \eta^{24}(p)$ the discriminant, a cusp form of weight
$12$ for $SL(2,\Z)$. The correction $p^{-\chi/24}$ has again an interpretation
as the regularized sum of zero-point energies. (Each boson contributes
$-1/24$, each fermion $+1/24$.)

\newsection{Light-Cone Quantization of String Theories}

It is now straightforward to generalize all this 
to string theory along the lines
\ban
\hbox{\it quantum mechanics on $SX$} & \ra &\hbox{\it quantum field theory
on $X$}\\
\hbox{\it 2-d conformal field theory on $SX$} & \ra  & \hbox{\it quantum string 
theory on $X$}
\ean
The interest in this generalization lies in particular in the absence of
a good Lorentz-invariant description of non-perturbative
second-quantized quantum string theory. So we can gain something by
studying the reformulation in terms of sigma models on symmetric products

It is not difficult to give a string theory interpretation of our
results in section 3 on sigma models on symmetric product spaces.
Clearly we want to identify the DLCQ of string theory on $(\R\times
S^1)^{1,1} \times X$ with the SCFT on $SX$. An obvious question is
which type of string theory are we discussing. Indeed, the number of
consistent interacting closed string theories is highly restricted:
the obvious candidates are

\begin{enumerate}

\item Type II and heterotic strings in 10 dimension.

\item Topological strings in all even dimensions.

\item Non-abelian strings in 6 dimension.

\end{enumerate}

Here the last example only recently emerged, and we will return to it in
section 7. We will start with the Type II string.

\newsubsection{The IIA superstring in light-cone gauge}

The physical states of the 
ten-dimensional type II superstring are most conveniently described
in the light-cone Green-Schwarz formalism. We usually think about
the superstring in
terms of maps of a Riemann surface $\Sigma$ into flat space-time
$\R^{1,9}$. But in light-cone gauge we make a decomposition 
$\R^{1,1} \times \R^8$ with
corresponding local coordinates $(x^+,x^-,x^i)$.  
The physical degrees of freedom are then completely
encoded in the transverse map
\be
x:\  \Sigma \ra \R^8.
\ee
The model has 16 supercharges (8 left-moving and 8 right-moving) and
carries a $Spin(8)$ R-symmetry.

More precisely, apart from the bosonic field $x$, we also have fermionic
fields that are defined for a general 8-dimensional transverse space $X$
as follows. Let $S^\pm$ denote the two inequivalent 8-dimensional spinor
representations of $Spin(8)$. We use the same notation to indicate the
corresponding spinor bundles of $X$. Let $V$ denote the vector
representation of $Spin(8)$ and let $TX$ be the associated tangent
bundle. In this notation we have
\be
\d x \in \G(K_\Sigma \otimes x^* TX),\qquad
\dbar x \in \G(\bar K_\Sigma \otimes x^* TX).
\ee
Now the left-moving and right-moving fermions
$\theta,\thetabar$ are sections of
\be
\th \in \G(K_\Sigma^{1/2} \otimes x^* S^+),\qquad
\bar\th \in \G(\bar K_\Sigma^{1/2} \otimes x^* S^\pm)
\ee
The choice of spin structure on $\Sigma$ is always Ramond or periodic.
The different choices of $Spin(8)$ representations for the right-moving
fermion $\bar\th$ ($S^+$ or $S^-$) give the distinction between the IIA
and IIB string. We will work with the IIA string for which the
representation of $\bar\th$ is chosen to be the conjugate spinor $S^-$,
but the IIB string follows the same pattern.

With these fields the action of the first-quantized sigma model is
simply the following free CFT
\be
S = \int d^2\!\s\; \Bigl(\half \d x^i \dbar x^i + \th^a \dbar \th^a + 
\bar\th^{\dot a} \d\bar\th^{\dot a}\Bigr).
\ee
This model has a Hilbert space that is of the form
\be
\cH = L^2(\R^8) \otimes \V \otimes \cF \otimes\bar\cF.
\ee
We recognize familiar components: the bosonic zero-mode space $L^2(\R^8)$
describes the quantum mechanics of the center of mass $\oint x^i$
of the string.
The fermionic zero-modes $\oint \th^a$, $\oint \bar\th^{\dot a}$
give rise to the  $16\times 16$ dimensional
vector space of ground states
\be
\V \cong (V \oplus S^-) \otimes  (V \oplus S^+)
\ee
where the spinor representations should be considered odd. This space
forms a representation of the Clifford algebra ${\rm Cliff}(S^+)\otimes
{\rm Cliff}(S^-)$ generated by the fermion zero mode
\be
\G^a = \int {d\s\over 2\pi}\; \th^a(\s),\qquad
\G^{\dot a} = \int {d\s\over 2\pi}\; \bar\th^{\dot a}(\s).
\ee
Using the triality $S^+\ra V \ra S^-\ra S^+$ of $Spin(8)$
this maps to the usual
Clifford representation of ${\rm Cliff}(V)$ on $S^+ \oplus S^-$.
Finally the Fock space $\cF$ of non-zero-modes is given by
\be
\cF_q = \bigotimes_{n>0} \Bigl( \ext_{q^n} S^- \otimes S_{q^n}V \Bigr)
\ee
with a similar expression for $\bar\cF$ with $S^-$ replaced by $S^+$.

In this light-cone gauge the coordinate $x^+$ is given by
\be
x^+(\s,\t) = p^+ \t
\ee
for fixed longitudinal momentum $p^+>0$, whereas $x^-$ is determined by the
constraints
\be
\d x^- = {1\over p^+} (\d x)^2,\qquad
\dbar x^- = {1\over p^+} (\dbar x)^2.
\label{x^-}
\ee
The Hilbert space of physical states of a single string
with longitudinal momentum
$p^+$ is given by the CFT Hilbert space $\cH$ restricted to states
with zero world-sheet momentum, the level-matching condition
\be
P= L_0 - \Lbar_0 =0.
\ee
The light-cone energy $p^-$ is then determined as by the
mass-shell relation
\be
p^- = {1\over p^+}(L_0 + \Lbar_0)= {1\over p^+}H.
\ee
One can also consider DLCQ with the null coordinate $x^-$ periodically
identified with radius $R$. This induced two effects. First, the
momentum $p^+$ is quantized as $p^+=n/R$, $n\in \Z_{>0}$. Second, it
allows the string to be wrapped around the compact null direction
giving it a non-trivial winding number 
\be
w^- = \int_{S^1} dx^- = 2\pi m R,\qquad m\in\Z.
\ee
However, using the constraints (\ref{x^-}) we find that
\be
w^- = {2\pi \over p^+}(L_0 - \Lbar_0) = {2\pi R \over n} (L_0 -
\Lbar_0).
\ee
So, in order for $m$ to be an integer, we see that the CFT Hilbert space
must now be restricted to the space $\cH^{(n)}$ consisting
of all states that satisfied the modified
level-matching condition
\be
P = L_0 - \Lbar_0 = \hbox{0 (mod $n$)}
\ee
This is exactly the defintion of the 
Hilbert space $\cH\nn$ in
section 3.5. (In the similar spirit the uncompactified model had
Hilbert space $\cH^{(\infty)}$.) This motivates us to describe
the second-quantized Type IIA string in terms of a SCFT on the
orbifold
\be
S^N\R^8 = \R^{8N}/S_N.
\ee
Indeed in this correspondence we have:
\ba
p^+ \is  N, \nonu
p^- \is H= L_0 + \Lbar_0,\\[2mm]
w^- \is P=L_0 - \Lbar_0.
\ea
This gives the following form for
the second-quantized Fock space
\be
\cF_p = \bigotimes_{n>0} S_{p^n}\cH\nn
\ee
where $p$ keeps again track of $p^+$. This is both the Hilbert space of the
free string theory and of the orbifold sigma model on $S_p\R^8$. So we can
identify their partition functions
\be
Z^{string}(\R^8;p,q,\qbar) = Z^{SCFT}(S_p\R^8;q,\qbar),
\ee
with
\ba
Z^{string}(\R^8;p,q,\qbar) \is  \Tr_{\strut \cF}
p^{P^+}q^{P^- + W^-}\qbar^{P^- - W^-},\nonu
Z^{SCFT}(S_p\R^8;q,\qbar) \is  \sum_{N\geq 0} p^N \Tr_{\strut \cH(S^N\R^8)}
q^{L_0-N/2}\qbar^{\Lbar_0-N/2},
\ea
(Here we used that the central charge is $12N$.)

Note that this sigma model is not precisely of the form as we discussed
in section 3. The world-sheet fermions now transform as spinors instead
of vectors of $Spin(8)$. The modification that one has to make are
however completely straightforward. In particular, the $U(1)$'s whose
quantum numbers gave the world-sheet fermion number $F_{L,R}$ only
emerge if we break $Spin(8)$ to $SU(4)\times U(1)\cong Spin(6)\times
Spin(2)$. 

This issue is directly related to compactifications. Can we 
consider for the transversal space instead of $\R^8$ a compact
Calabi-Yau  four-fold $X$ and make contact with our computations of
the elliptic genus of $SX$? 
In the non-linear sigma models of section 3 the fermion
fields where always assumed to take values in the (pull-back of the)
tangent bundle to the target space $X$, wheras in the Green-Schwarz
string they are sections of spinor bundles. Note however that on a
Calabi-Yau four-fold we have a reduction of the structure group $SO(8)$
to $SU(4)$.
Under this reduction we have the following well-known
decomposition of the three
8-dimensional representations of $Spin(8)$ in terms of the
representations of $SU(4) \times U(1)$ (up to triality)
\ba
V & \ra & {\bf 4}_{1/2} \oplus \bar{\bf 4}_{-1/2} \nonu
S^+ & \ra & {\bf 4}_{-1/2} \oplus \bar{\bf 4}_{1/2}, \\[2mm]
S^- & \ra & {\bf 1}_1 \oplus {\bf 6}_0 \oplus {\bf 1}_{-1}.\nonumber
\ea
So we see that, as far as the $SU(4)$ symmetry is concerned,
if we only use the $S^+$ representation, we could just
as well worked with the standard $N=2$ SCFT sigma-model, since this
spinor bundle
is isomorphic to the tangent bundle. 

This remark picks out naturally the Type IIB string, whose world-sheet
fermions carry only one $Spin(8)$ chirality that we can choose to be
$S^+$. So in this formulation only the type IIB light-cone model allows
for a full compactifications on a Calabi-Yau four-fold. This fact is
actually well-known. The IIA string acquires an anomaly $\chi(X)/24$
that has to be cancelled by including some net number of strings
\cite{svw,mukhi}. For the Type IIB string this translates under a
T-duality to a net momentum in the vacuum; we will see this fact again
in a moment. We can also work with only left-moving BPS strings that are
related to the elliptic genus of the SCFT. In that case it does not
matter if we choose the IIA or IIB strings. 

\newsubsection{Elliptic genera and automorphic forms}

If we just want to discuss free strings, without interactions, say in
light-cone gauge without insisting on Lorentz-invariance, there are many
more possible strings than the ten-dimensional superstring. 
In particular we can consider a string whose
transverse degrees of freedom are described by the $N=2$ supersymmetric
sigma model on the Calabi-Yau space $X$. This string has as its
low-energy, massless spectrum the field theory that consists of all
$k$-form gauge fields, that we discussed in section 4.3. This is by the
way exactly the field content of the topological string, that can be
defined in any (even) dimension, but which only has non-vanishing
interactions without gravitational descendents in space-time dimension
6. So this critical case corresponds to choosing a transversal four-fold
or complex surface $X$. If $X$ has to be compact that restricts us to
$T^4$ or $K3$. We will return to this topic in section 7.

Therefore another class of free string theories to be considered in the
light-cone formulation are the `untwisted' versions of the topological
string, where we do not  impose the usual BRST cohomology $Q_L=Q_R=0$ that
reduces the string to its massless fields. In fact, another interesting case is
the half-twisted string (see section 3.3) in which we only impose
$Q_R=0$. For that model we expect to make contact with the elliptic
genus.

Indeed, in that case there is a straightforward
explanation of the automorphic properties of the elliptic genus
of the symmetric product. 
We recall the main formula (\ref{genus}) of Theorem 2, that we now
interpret as a partition function of second-quantized BPS strings
\be
Z^{string}(X;p,q,y) = \chi_{orb}(S_pX;q,y) = 
\prod_{n>0,\,m\geq 0,\,\ell} (1-p^nq^m y^\ell)^{-c(nm,\ell)}
\ee
with the coefficients $c(m,\ell)$ determined by the elliptic genus of
$X$, 
\be
\chi(X;q,y) = \sum_{m,\ell} c(m,\ell) q^m y^\ell.
\ee
Note that since the strings carry only left-moving excitations,
$\Lbar_0=0$, the space-time Hamiltonian $p^-$ and winding number $w^-$
can be identified and thus the partition function represents the
{\it space-time} character
\be
Z^{string}(X;p,q,y) = \Tr_{\strut \cF} (-1)^F p^{P^+} q^{W^-} y^F.
\ee
This is precisely the object we promised in our discussion to study.

We will parametrize $p,q,y$ as
\be
p=e^{2\pi i \s},\ q = e^{2\pi i \t},\ y ^{2\pi i z},
\ee
or equivalently by a $2\times 2$ period matrix
\be
\W = 
\twomatrixd \s z z \t
\ee
in the Siegel upper half-space, $\det\Im\W > 0$.
The group $Sp(4,\Z) \cong SO(3,2,\Z)$ acts on the matrix $\W$ by
fractional linear transformations, $\W \ra (A\w + B)(C\W + D)\inv$.

Now the claim is that the string partition function $\chi_{orb}(X:p,q,y)$ is
almost equal to an automorphic form for the group $SO(3,2,\Z)$, of
the infinite product type as appear in the work of Borcherds
\cite{borcherds}. This  is just the string theory generalization of
the fact that the Euler number $\chi_{orb}(S_pX)$ is almost a modular form
of $SL(2,\Z)$. In fact, the Euler number is obtained from the
elliptic genus in the limit
$y\ra 1$, $z\ra 0$, where the $q$-dependence disappears. 
In this case, the automorphic group degenerates as
\be
Sp(4,\Z) \ra SL(2,\Z) \times SL(2,\Z)
\ee
where only the first $SL(2,\Z)$ factor acts non-trivially on $p$.
 
The precise form of the corrections needed to get a true
automorphic function $\F(p,q,y)$ for a general Calabi-Yau $d$-fold $X$ has 
been worked out in detail in \cite{neumann}. It is defined by the product
\be
\Phi(p,q,y) = p^a q^b y^c \prod_{(n,m,\ell)>0}(1-p^n q^m 
y^\ell)^{c(nm,\ell)}
\ee
where the positivity condition means: $n,m\geq0$ with $\ell>0$ in the case 
$n=m=0$. The `Weyl vector' $(a,b,c)$ is defined by
\be
a=b= \chi(X)/24,\qquad c=\sum_\ell -{|\ell|\over 4} c(0,\ell).
\label{weyl}
\ee
Here the coefficients $c(0,\ell)$ are the partial Euler numbers
\be
c(0,r-{d\over 2}) = \sum_s (-1)^{s+r}h^{r,s}
\ee
One can then show that $\Phi$ is an automorphic form of
weight $c(0,0)/2$ for the group $O(3,2,\Z)$ for a suitable quadratic
form of signature $(3,2)$.

The form $\Phi$ follows actually from a standard one-loop string
amplitude defined as an integral over the fundamental domain
\cite{harvey-moore,kawai}.  The integrand consists of the genus one
partition function of the string on $X\times T^2$ and has a manifest
$SO(3,2,\Z)$ T-duality invariance. The $SO(3,2,\Z)$ appears in the 
following way. First of all, as explained in the introduction, strings
on $T^2$ have two quantized momenta and two winding numbers, giving the
Narain lattice $\G^{2,2}$. For a transversal Calabi-Yau space, there are
also the left-moving and right-moving Fermi numbers $F_L,F_R$. Since we
restrict to right-moving ground states in the elliptic genus, only $F_L$
gives another integer conserved quantum number $\ell$. Adding this charge to
the Narain lattice enlarges it to $\G^{3,2}$. Moreover, it allows us to 
extend the moduli $\s,\t$ of the two-torus by another complex parameter 
$z$ that couples to  $F_L=\ell$. Technically, $z$
has an interpretation as a Wilson loop that
parametrizes the $U(1)_L$ bundle over $T^2$.  Together, $\s$, $\t$, $z$
parametrize the lattice $\Gamma^{3,2}$; they can be considered as a point
on the symmetric space
\be
SO(3,2)/SO(3)\times SO(2) \cong \H^{2,1}.
\ee
Now the strategy is to compute the string partition function
through a one-loop amplitude
\be
Z^{string}(X;p,q,y) = \exp F^{string}(X;p,q,y).
\ee
Note that $F^{string}$ is the partition function for maps from the
world-sheet elliptic curve, with a modulus that we denote as $\t'$, 
to the space-time that contains an elliptic curve with modulus $\t$,
\be
T^2_{\t'} \ra T^2_{\s,\t,z}\times X.
\ee
It is easily to confuse the two elliptic curves!
One now computes an integral over the fundamental domain of the
world-sheet modulus $\t'$ that has the form
\be 
F^{string} = {\textstyle 1\over 2} \int {d^2\tau'\over
\tau'_2} \sum_{-{d\over 2} +1 \leq \e \leq {d\over 2}}
\sum_{\mbox{\footnotesize $(p_L,p_R)\in\Gamma_\e^{3,2}$} \atop
\mbox{\footnotesize $n \in 2d\Z - \e^2$}} e^{i\pi(\tau' p_L^2-
\overline\tau' p_R^2)} c_\e(n) e^{\pi i n\tau'/d}
\label{int}
\ee 
where the notation $\G^{3,2}_\e$ indicates that $\ell=\e$ (mod $2d$)
and where the coefficients $c_\e(n)$ are defined in terms of the 
expansion coefficients  of the elliptic genus of $X$ as
$c(m,\ell) = c_\e(2dm-\ell^2),$ with $\e=\ll$ (mod $2d$).
This integral can be computed using
the by-now standard techniques of \cite{dixon,harvey-moore,kawai,kawai-2}.
The final result of the integration is \cite{neumann},
\be
F^{string}(\W,\bar\W) = -\log \left((\det\Im\W)^{c(0,0)/2} |\Phi(\W)|^2\right)
\ee
Since the integral $F$ is by construction invariant under the T-duality
group $O(3,2,\Z)$, this determines the automorphic properties of $\Phi$.
The factor $\det\Im\W$ transforms with weight $-1$, which fixes the
weight of the form $\Phi$ to be $c(0,0)/2$. 
This formula should be contrasted with the analogous computation for
the zero-modes (\ref{par}), \ie, the field theory limit,
\be
F^{QFT}(\t,\bar\t) = - 
\log \left((\Im\t)^{c/2} |\eta^c(\t)|^{2}\right),\qquad
c=\chi(X).
\ee

In the special case of $K3$ the infinite product $\F(\W)$ is a
well-known automorphic form \cite{gritsenko}, see also \cite{gn,ff}.
First of all, the elliptic genus of $K3$ is the unique (up to a scalar)
weak Jacobi form of weight 0 and index 1. Realizing $K3$ as a Kummer
surface (resolving the orbifold $T^4/\Z_2$) we see that the elliptic
genus can be written in terms of genus one theta-functions as
\be
\chi(K3;p,q) = 2^3 \sum_{even\ \a} {\vartheta^2_\a(z;\t) \over
\vartheta^2_a(0,\t)}.
\ee
If we now identify $\W$ with the period matrix of a
genus two Riemann surface, we can rewrite the automorphic form
in terms of genus two theta-functions,
\be
\F(\W) = 2^{-12} \prod_{even\ \a} \vartheta[\a](\W)^2
\ee
In the work of Gritsenko and Nikulin \cite{gritsenko} is is shown that
$\F$ also has an interpretation as the denominator of a generalized
Kac-Moody algebra. It is a rather obvious conjecture that this GKM
should be given by the algebra of BPS states induced by the
string interaction. The full story for $K3$ is quite beautiful
and explained in \cite{kawai-2}. 
See \cite{harvey-moore-2,dimitri} for more on the
connection with GKM's.

Summarizing we have seen the following:

\begin{enumerate}

\item The (BPS) string theory partition function factorizes in left-moving
and right-moving contributions that are holomorphic functions of the
moduli $p,q,y$. 

\item The holomorphic contributions are automorpic forms of weight
$-c(0,0)/2$ of the group $SO(3,2,\Z)$ if a particular correction is added.
This correction takes the form \be (pq)^{\chi(X)/24}y^c
\prod_{\ell>0}(1-y^\ell)^{c(0,\ell)}
\prod_{m>0,\ell}(1-q^my^\ell)^{c(m,\ell)} \ee These three factors have
the following interpretation. The first factor is again the regulated
zero-point energy, very similar to the field theory result. The second
factor is due to the bosonic and fermionic zero-modes.
(Recall that the low-energy field theory describes general differential
forms on $T^2 \times X$.) The third factor is there to restore the
symmetry in $p$ and $q$. It can only be understood using $T$-duality.

\item The full partition function is invariant because the zero-mode
contribution adds a non-holomorphic factor $(\det\Im\W)^{-c(0,0)/2}$.

\item The holomorphic contributions are characters of an
infinite-dimensional generalized Kac-Moody algebra, directly related to
the creation and annihilation operators of the string Fock space and
their interactions.

\item The modularity of the characters, \ie, the transformation
properties under the automorphic group $SO(3,2,\Z)$ is `explained' by the
relation to a partition function of a string on a two-torus $T^2$
with an associated line bundle with moduli $\tau,\s,z$ 
and $T$-duality group $SO(3,2,\Z)$.

\end{enumerate}

\newsection{Matrix Strings and Interactions}

Up to now we have only considered free theories and observed how in
light-cone quantization these models could be reformulated using
first-quantized theories on symmetric products. Now we want to atke
advantage from this relation to include interactions. This has proven
possible for two important examples: 1) the ten-dimensional IIA
superstring and some of its compactifications, and 2) the class of (2,0)
supersymmetric six-dimensional non-abelian string theories. By taking
the low-energy limit, similar formulations for the field theory limits
follow. The essential starting point in these constructions is the
beautiful Ansatz for a non-perturbative formulation of M-theory known as
matrix theory \cite{bfss}. See for example the reviews
\cite{susskind,banks,maxmic} for more information about matrix theory.

\newsubsection{Supersymmetric Yang-Mills theory}

Matrix string theory gives a very simple Ansatz of what
non-perturbative IIA string theory looks like in light-cone gauge
\cite{motl,banks-seiberg,matrix-string}. It is
simply given by the maximally supersymmetric two-dimensional
Yang-Mills theory with gauge group $U(N)$ in the limit $N\ra \infty$
(or with finite $N$ in DLCQ).

To be more precise, let us consider two-dimensional $U(N)$
SYM theory with 16 supercharges.
It can be obtained by dimensionally reducing the $\cN=1$ SYM theory in
10 dimensions. Its field content consists of the following fields.
First we pick a (necessarily trivial) $U(N)$ principle bundle $P$ on
the world-sheet $S^1 \times \R$. Let $A$ be a connection on this
bundle.  We further have 8 scalar field $X^i$ in the vector
representation $V$ of $Spin(8)$, and 8 left-moving fermions $\th^a$ in 
the spinor representation $S^+$
and 8 right-moving fermion $\bar\th^{\dot a}$ in the conjugated spinor
representation $S^-$. All these
fields are Hermitean $N\times N$ matrices, or if one wishes sections
of the adjoint bundle ${\rm ad}(P)$ .

The action for the SYM theory reads
\ba
S_{SYM} \is \int d^2\s\; \Tr\Bigl(\half |DX^i|^2 + 
\th^a \bar D \th^a + \bar\th^{\dot a} D \bar\th^{\dot a} \nonu
&&\qquad + {1\over 2g^2} |F_A |^2 + g^2 \sum_{i<j} [X^i,X^j]^2 + g \,
\bar\th^{\dot a}\g^i_{a\dot a}[X^i,\th^a]\Bigr)
\label{sym-action}
\ea
Here $g$ is the SYM coupling constant ---- a dimensionful quantity with
dimension 1/length in two dimensions. This means in particular that
the SYM model is not conformal invariant. In fact, at large length
scales (in the IR) the model becomes strongly interacting. So we have
a one-parameter family of QFT's labeled by the coupling constant $g$
or equivalently a length scale $\ell=1/g$.

The relation with string theory is the following. First of all for finite $N$
the
Hilbert space of states of the SYM theory should be identified with
the DLCQ {\it second-quantized} IIA string Hilbert space.  The integer $N$
that gives the rank of the gauge group
is then related to the total longitudinal momentum in the usual way as
\be
p^+ = N/R,
\ee
whereas the total light-cone energy is given by
\be
p^- = {N\over p^+}H_{SYM}
\ee
with $H_{SYM}$ the Hamiltonian of the SYM model. Note that in the
decompactification of the null circle where we will take
$N,R\ra \infty$, keeping their ratio finite, only SYM states with energy
\be
H_{SYM} \sim {1\over N}
\ee
will contribute a finite amount to $p^-$. Finally, the IIA string
coupling constant $g_s$ (a dimensionless constant) is identified as
\be
g_s = (g\ell_s)\inv
\ee
with $\ell_s$ the string length, $\a'=\ell_s^2$.

From this identification we see that free string theory ($g_s=0$) is
recovered at strong SYM coupling ($g=\infty$). This is equivalent to the
statement that free string theory is obtained in the IR limit. In this
scaling limit---the fixed point of the renormalization group flow---we
expect on general grounds to recover a superconformal field theory with
16 supercharges. We will now argue that this SCFT is the supersymmetric
sigma model with target space $S^N\R^8$. We can then use our previous
analysis of orbifold sigma models to conclude that the point $g_s=0$
indeed describes the second-quantized free IIA string.

The analysis proceeds in two steps. First we observe that because of the
last two terms in the action (\ref{sym-action}), in the limit $g_s=0$
which is equivalent to $g=\infty$, the fields $X$ and $\theta$
necessarily have to commute. This means that we can write the matrix
coordinates as
\be
X^i(\s) =  U(\s) \. x^i(\s) \. U^{-1}(\s),
\ee
with $U\in U(N)$ and $x^i$ a diagonal matrix with eigenvalues
$x_1^i,\ldots, x_N^i$. Now the matrix valued fields $X^i(\s)$ 
are single-valued,
being section of the trivial bundle $U(N)$ vector ${\rm ad}(P)$.
But this does not imply that the fields $U(\s)$ and $x^i(\s)$ are too.
In fact, it is possible that after a shift $\s \ra \s+2\pi$ the
individual eigenvalues are permuted due to a spectral flow. Only the
set of eigenvalues (or more properly the set of common eigenstates) of
the commuting matrices $X^i$ is a gauge invariant quantity. So we
should allow for configurations of the form
\be
x^i(\sigma + 2\pi) = g \. x^i(\sigma) \. g^{-1},
\ee
with $g\in S_N$ the Weyl group of $U(N)$. Effectively this tells us
that we are dealing with an orbifold with target space 
\be
\R^{8N}/S_N = S^N\R^8,
\ee
given Lie-theoretically as $t^8/W$ with $t$ the Cartan Lie algebra and
$W$ the Weyl group of $U(N)$. 

As we have analyzed before this implies
that the Hilbert space decomposes in superselection sectors labeled by
the conjugacy classes $[g]$ of $S_N$, which in turn are given by
partitions of $N$. This structure indicates that the Hilbert space 
is a Fock space of second-quantized IIA strings. A sector twisted by
\be
g = (n_1)\ldots(n_k)
\ee
describes $k$ strings of longitudinal momentum
\be
p^+_i = {n_i\over R} = {n_i\over N} p^+_{tot},\qquad i=1,\ldots,k.
\ee
We have also seen how for a string with a twist $(n)$ of `length' $n$
the $Z_n$ projection of the orbifold projects the Hilbert space to
a subsector conditioned to
\be
L_0 - \Lbar_0 = \hbox{0 (mod $n$)}
\ee
that we now interpreted as the usual DLCQ level-matching condition.
In the large $N$ limit, also the individual $n_i$ go to infinity,
effectively decompactifying the null circle.

The second step consists of analyzing the behaviour of the gauge
field. The possibly twisted configurations of $X^i(\s)$ break the gauge
group $U(N)$ to an abelian subgroup $T$ that commutes with the
configuration $X^i(\s)$. In fact, if the twist sector is labeled by a
partition
\be
n_1 + \ldots + n_k = N
\ee
describing $k$ strings of length $n_1,\ldots,n_k$, the unbroken gauge
group is 
\be
T \cong U(1)^k.
\ee 
Because of the Higgs effect all the broken components of the gauge field
acquire masses of the order $g$ and thus decouple in the IR limit. This
leaves us with a free abelian gauge theory on $\R \times S^1$. This
model has been analyzed in great detail. Dividing by the gauge
symmetries leaves us with the holonomy along the $S^1$
\be
Hol(A) = e^{\oint_{S^1} A} \in T
\ee
as the only physical degree of freedom. The gauge theory is
therefore described by the quantum
mechanics on the torus $T$ with Hamiltonian given by
\be
H = - g^2 \Delta
\ee
with $\Delta $ the Laplacian on $T$. The eigenstates are given by the
characters of the irreducible representations of $T$ with eigenvalues
(energies) $g^2$ times the second Casimir invariant of the
representation. Clearly in the limit $g\ra \infty$ only the vacuum state
or trivial representation survives. This state has a constant
wavefunction on $T$ which has the interpretation that the abelian gauge
field is free to fluctuate, a result from the fact that in strong
coupling the action $S={1\over g^2}\int F^2$ goes to zero. So all-in-all
the gauge field sector only contributes a single vacuum state. This
completes our heuristic derivation of the IR limit of SYM. 

Since two-dimensional gauge theories are so well-behaved it would be
interesting to make the above in a completely rigorous statement about
the IR fixed point of SYM. One of the points of concern could be
complications that emerge if some of the eigenvalues coincide. In that
case unbroken non-abelian symmetries appear. As we will show in the next
section however, from the SCFT perspective such effects are always
irrelevant and thus disappear in the IR limit. In fcat, these effects are
exactly responsible for the perturbative interactions at finite $g$.

\newsubsection{Interactions}

If the matrix string theory conjecture is correct, for finite coupling
constant the SYM theory should reproduce the interacting string. A
non-trivial check of this conjecture is to identify the 
correction for small $g_s$. This should be given by the joining and
splitting interaction of the strings, producing surfaces with
nontrivial topology.

This computation was done in \cite{matrix-string} where the leading correction
was computed. Let us try to summarize this computation. 
(It is also reviewed in \cite{maxmic}.) The idea is to
analyze the behaviour of the SYM theory in the neighbourhood of the IR
fixed point. In leading order, a deformation to finite $g$, is given
by the least irrelevant operator in the orbifold CFT. That is, we look
for the operator $\cO$ in the sigma model that preserves all
the supersymmetries and the $Spin(8)$ R-symmetry
and that has the smallest scaling dimensions. The deformed QFT
then has an action of the form
\be
S = S_{SCFT} + (g_s)^{h-2} \int \cO + \ldots
\ee
with $h$ the toal scaling dimension of $\cO$. We would like to see
that the power of $g_s$ is one (so that $h=3$) and that this deformation
induces the usual joining and splitting interaction.

Note that the Hilbert space of the matrix string was defined with Ramond boundary
conditions for the supercurrent $G^{\dot a} = \g^{a\dot a}_i \th^a \d
x^i.$ That is, we have
\be
G^{\dot a}(\s + 2\pi) = G^{\dot a}(\s).
\ee
We have seen that the ground state space $\V\nn$ of a $\Z_n$ twisted
sector $\cH\nn$ is isomorphic to the ground state space of a single
string
\be
\V\nn \cong (V \oplus S^-) \otimes  (V \oplus S^+).
\ee
Only the conformal dimensions are rescaled and given by
\be
L_0  = \Lbar_0 = n d/8,
\ee
since the central charge of the SCFT is $n$ times as big. Here $d$ was
the complex dimension of the target space, so in our case
$d=4$. 

One way to understand this vacuum degeneracy is that $Z_n$ action on the
$n$ fermions $\th_1,\ldots,\th_n$ can be diagonalized with
eigenvalues $e^{2\pi ik/n}$, $k=0,\ldots,n-1$. That is, there are linear
combinations of the $\th_k$, let us denote them by $\tilde \th_k$
that have boundary conditions
\be
\tilde \th_k(\s + 2\pi) = e^{2\pi i k\over n} \tilde\th_k(\s).
\ee
So the linear sum
\be
\tilde\th_0 = \th_1 + \ldots + \th_n
\ee
is always periodic and its zero modes give the 16 fold vacuum degeneracy.
A similar story holds for the right-moving fermions.

Since we want to keep Ramond boundary conditions in the interacting
theory, the local operator $\cO$ that describes the first-order
deformation should be in the NS-sector. This just tells us that
the OPE
\be
G^{\dot a}(z) \cO(w)
\ee
is single-valued in $z-w$. So, using the familiar operator-state
correspondence of CFT we have to look in the NS-sector of the Hilbert
space. These are of course again labeled by twist fields. The only
difference is that the fermions now have an extra minus sign in their
monodromy, and satisfy the boundary conditions
\be
\tilde \th_k(\s + 2\pi) = - e^{2\pi i k\over n} \tilde\th_k.
\ee
Now depending on whether $n$ is even or odd there is a periodic
fermion or not. So we expect to find only a degeneracy for even $n$.
It is not difficult to compute the conformal dimension
of the NS ground state in a $Z_n$ twisted sector. First of all, both
for the bosons and fermions the $Z_n$ action can be diagonalized.
The bosonic twist field that implements a twist with eigenvalue
$e^{2\pi i k/n}$ has conformal dimensions $dk(n-k)/2n^2$, with
$d$ the complex dimension of the transversal space ($d=4$ for the
IIA string). For the
corresponding fermionic twist field we find conformal dimension
$dm^2/2n^2$, where $m={\rm min}(k,N-k)$. 
Adding up all the possible eigenvalues we obtain
total conformal dimension 
\be
h = \option{n}{$n$ even}{n-{1\over n}}{$n$ odd}
\ee
In particular the 
lowest dimension $h=2$ is given by the $Z_2$ twist field $\s$. Since
$n=2$ is even, this ground state has the usual degeneracy
\be
\s \in (V \oplus S^-) \otimes  (V \oplus S^+)
\ee
Note that the zero-modes of the superpartner of the twisted boson $x^i$
give this degeneracy. However, the NS ground state
is not supersymmetric neither $Spin(8)$ invariant, and is therefore not a
suitable candidate for our operator $\cO$.

There is however a small modification that does respect the supersymmetry
algebra. In the $Z_2$ twisted sector the coordinate $x^i$ has a mode
expansion
\be
\d x^i = \sum_{n\in \Z+{1\over 2}} \a^i_n z^{-n-1}.
\ee
We now consider the first excited state
\be
\cO = \a^i_{-1/2}\bar \a^j_{-1/2} \s^{ij}
\ee
of conformal weights $2+1=3$. (Here $\s^{ij}$ indicates the components
of $\s$ in $V\otimes V$.
This operator can be written as
\be
\cO = G^{\dot a}_{-1/2}\bar G^{b}_{-1/2}\s^{\dot a b}
\ee
and therefore satisfies
\be
[G^{\dot a}_{-1/2},\cO] = \d \bar G^{b}_{-1/2}\s^{\dot a b}
\ee
which is sufficient. Since $\cO$ is both SUSY and $Spin(*)$ invariant,
it is the leading irrelevant operator that we were looking for.

What is the interpretation of the field $\cO$ in string perturbation
theory? It clearly maps superselection sectors with two strings into
sectors with one string and vice versa. It is therefore exactly the
usual joining and splitting interaction. In fact, the perturbation in the
operator $\cO$ reproduces the standard light-cone perturbation theory.

There is also a clear geometric interpretation of the twist field
interaction $\cO$.  Consider the manifold $\R^8/\Z_2$ or if one wishes
the compact version $T^8/Z_2$. This is a Calabi-Yau orbifold and
defines a perfectly well-behaved superconformal sigma model.
One could now try to blow-up
the $Z_2$ singularity to obtain a smooth Calabi-Yau space. 
It is well-known that this cannot be done
without destroying the Calabi-Yau property;
the orbifold $\R^8/\Z_2$ is rigid. In the SCFT language this is
expressed by the fact that corresponding deformation does not respect
the superconformal algebra. Algebraically, preserving the conformal
invariance implies that the operator is marginal with scaling dimension
2. The fact that we found weight $3$ is therefore in accordance with
the fact that the two-dimensional field theory deforms to a massive
field theory with a length scale --- two-dimensional SYM.

However, we see that if the transverse target space would have been
four-dimensional, the twist field interaction would have
$L_0=\Lbar_0=1$ and would have represented a marginal operator. This
is a simple reflection of the fact that the orbifold $\R^4/\Z_2$ or
$T^4/\Z_2$ can be resolved to a smooth Calabi-yau manifold,
respectively an hyperk\"ahler ALE space or a K3 surface. We therefore
turn now to the case where this four-dimensional example becomes
relevant.

\newsection{String Theories in Six Dimensions}

For superstrings the critical dimension is ten, or transversal dimension
eight. However, in the past year that has been growing evidence that
there is also a fascinating class of string theories with critical
dimension six, with a four-dimensional transversal space. In fact, there
is believed to be such a string for every simply-laced Lie group of type
$A,D,E$. We will mainly focus on the $U(k)$ case.

There is very little know about these theories
\cite{witten-20,seiberg-20,seiberg-notes}. They have $(2,0)$
supersymmetry with a $Spin(4)$ R-symmetry, do not contain a graviton,
and give non-trivial six-dimensional SCFT's in the IR, where the
R-symmetry is enlarged to $Spin(5)=Sp(2)$. Roughly their massless modes
should be a theory of non-abelian two-form gauge fields, whose
three-form field strength is self-dual. For the $U(1)$ theory this can
be made precise. The massless modes form the irreducible (2,0) tensor
multiplet which consists of one two-form $B$ together with five scalar
field $X^I$.

Furthermore the string coupling of these microstrings or little strings
is believed to be fixed to one (basically because of the self-duality).
Since the string coupling cannot be tuned to zero, there is no reason
why a free string spectrum should emerge. This good, because we know
that the six-dimensional Green-Schwarz superstring is not Lorentz
invariant. It is true however that this string does reproduce the tensor
multiplet as its massless sector.

\newsubsection{DLCQ formulations and matrix models}

Up to now we only know how to describe these (2,0) strings in a matrix
theory DLCQ formulation \cite{5d,abkss,witten-higgs,abs}. We choose a
six-dimensional space-time of the form
\be
M^{1,5} = (\R\times S^1)^{1,1} \times X^4
\ee
with $X$ a (Ricci flat) Riemannian four-manifold. We will often choose $X$
to be compact, which restricts us to either $T^4$ or $K3$.
We fix the longitudinal momentum to
be $p^+ = N/R$, with $R$ the radius of the null-circle. The claim is now
that this string theory can be described in terms of a two-dimensional 
sigma model with target space the moduli space 
\be
\cM_{k,N}(X) \nonumber
\ee
of $U(k)$ instantons (self-dual connections) on $X$ with
total instanton charge $ch_2=N$. This moduli space is a hyperk\"ahler 
manifold of real dimension $4Nk$. It has singularities,
corresponding to (colliding) point-like instantons. There is however a
particularly nice compactification by considering the moduli space
\be
\bar\cM_{k,N}(X) \nonumber
\ee
of (equivalence classes) of coherent torsion free sheaves of rank $k$ and 
$ch_2=N$. In particular for the case $k=1$ we find in this way the Hilbert 
scheme of dimension zero subschemes of length $N$
\be
\bar \cM_{1,N}(X) = \Hilb^N(X). 
\ee
This space is a intricate smooth resolution of the symmetric space $S^NX$
 \cite{nakajima}.
The fibers of the projection $\Hilb^N(X) \ra S^NX$ over the various
diagonals keep track of the particular way the points approach each
other. Quite generally, if $X$ is a smooth Calabi-Yau space of complex
dimension $d$ then the symmetric product $\SNX$ is also Calabi-Yau
manifold,
albeit an orbifold, now of dimension $Nd$. Only for (complex)
dimension two, \ie, if $X$ a four-torus or $K3$ surface, is it
possible to resolve the singularities of $\SNX$ to produce smooth
Calabi-Yau. The Hilbert scheme $\Hilb^N(X)$ provides a canonical
construction. For CY $d$-folds with $d\geq 3$ the Hilbert scheme is not
smooth.

We should mention here that all of the spaces $\cM_{k,N}$ are 
to be hyperk\"ahler deformations of $S^{Nk}X$. In
particular this implies that their cohomology is given by that of
the symmetric product. For more on this issue see \cite{inst}.

\newsubsection{Deformations and interactions}

For any Calabi-Yau space $X$, its deformation space is locally given
by $H^1(T_X)\cong H^{1,d-1}(X)^*$.  By a well-know result of Tian
and Todorov there are no obstructions to such deformations, and
therefore the dimension of the moduli space $\cM_X$ of inequivalent
complex structures is given by
$h^{1,d-1}(X)$. It is not difficult to compute the dimension of the
deformation space of the symmetric product $\SNX$ using the above
formalism. We see that there is always a contribution given by
$H^1(T_X)$. This corresponds to simply deforming the underlying manifold
$X$. However, for dimension $d=2$ and only for this dimension, there is
a second contribution coming from $H^0(X^{(2)})$. In fact, for $d=2$ we
have
\be
\dim \cM_{\SNX} = \dim \cM_X + 1.
\ee
There is a direct geometric interpretation of this extra deformation.
$X^{(2)}$ represents the small diagonal in $X^N$ where two points
coincide.  In the orbifold cohomology of $\SNX$ it contributes the
cohomology of $X$, shifted however in bi-degree by
$(d-1,d-1)=(1,1)$. The corresponding deformation corresponds to
blowing up in the given complex structure this small diagonal.

The corresponding operator in the SCFT is exactly the same $\Z_2$ twist
field that we have discussed before for the type II string. Therefore
this deformation can be given an interpretation as tuning the string
coupling constant \cite{inst}.

\vspace{8mm}

{\noindent \bf Acknowledgements}

\vspace{2mm}

A very much shortened version of these notes can be found in
\cite{icm}. I wish to thank the organisers of the {\it Geometry and
Duality Workshop} at the Institute for Theoretical Physics, UC Santa
Barbara, January 1998 and the {\it Spring School on String Theory and
Mathematics,} Harvard University, May 1998 for the invitation to
present these lectures.

\renewcommand{\Large}{\normalsize}
\renewcommand{\tt}{}

\end{document}